\documentclass[journal,twoside,web]{util/ieeecolor}

\makeatletter
\def\endthebibliography{%
	\def\@noitemerr{\@latex@warning{Empty `thebibliography' environment}}%
	\endlist
}
\makeatother

\usepackage{util/generic}
\usepackage{amsfonts}
\usepackage{amsmath}
\usepackage{graphicx}
\usepackage{subcaption}
\usepackage[ruled]{algorithm2e}
\usepackage{verbatim}
\usepackage{amssymb}
\usepackage{amsthm}
\usepackage{fancyhdr}

\DeclareMathOperator*{\argmax}{arg\,max}

\allowdisplaybreaks

\newtheorem{theorem}{Theorem}
\newtheorem{lemma}{Lemma}
\newtheorem*{example*}{Example}
\newtheorem{definition}{Definition}

\newcommand{\ag}{{\rm ag}}

\newcommand{\nog}{{\rm ng}}

\newcommand{\fim}{{\rm FIM}}

\def\BibTeX{{\rm B\kern-.05em{\sc i\kern-.025em b}\kern-.08em
		T\kern-.1667em\lower.7ex\hbox{E}\kern-.125emX}}
\title{The Impact of Measurement Passing in Sensor Network Measurement Selection}
\author{Authors}

\author{David Grimsman, Matthew R. Kirchner, Jo\~{a}o P. Hespanha and Jason R. Marden
\thanks{This paper was submitted for review on September 23, 2020: a preliminary version appears in \cite{Grimsman2020}, however many changes and result extensions separate these two works. This research was supported in part by AFRL contracts FA8750-18-C-0014 and  FA95550-20-1-0054 and ONR grants N00014-20-1-2359 and N00014-20-1-2093. }
\thanks{D. Grimsman (\texttt{grimsman@cs.byu.edu}) is with the Computer Science Department at Brigham Young University, Provo, UT}
\thanks{M. R. Kirchner (\texttt{kirchner@ucsb.edu}), J. R. Marden (\texttt{jmarden@ece.ucsb.edu}), and J. P. Hespanha (\texttt{hespanha@ece.ucsb.edu}) are with the Department of Electrical and Computer Engineering, University of California, Santa Barbara, CA}
}

\begin{document}

\maketitle

\begin{abstract}
	This paper considers a set of sensors, which as a group are tasked with taking measurements of the environment and sending a small subset of the measurements to a centralized data fusion center, where the measurements will be used to estimate the overall state of the environment. The sensors' goal is to send the most informative set of measurements so that the estimate is as accurate as possible. This problem is formulated as a submodular maximization problem, for which there exists a well-studied greedy algorithm, where each sensor sequentially chooses a set of measurements from its own local set, and communicates its decision to the future sensors in the sequence. In this work, sensors can additionally share measurements with one another, in order to augment the decision set of each sensor. We explore how this increase in communication can be exploited to improve the results of the nominal greedy algorithm. Specifically, we show that this measurement passing can improve the quality of the resulting measurement set by up to a factor of $n+1$, where $n$ is the number of sensors.
\end{abstract}
\section{Introduction}

Multiagent robotic systems have received increased attention in recent years. Research has addressed a large variety of challenges posed by such systems, for instance collaborative motion planning \cite{stipanovic2004decentralized,debord2018trajectory} and allocation \cite{choi2009consensus}. One area of importance is the coordination of vehicles by sharing information on a resource constrained communication network. This paper studies a system of heterogeneous sensors collecting data at high rate. Each sensor is connected to an off-site data fusion center with limited bandwidth, and the group of sensors is tasked with sending a small but high-quality set of measurements to the data fusion center for processing. In addition, the sensors can coordinate which measurements to send via a local communication network. This problem has been studied in \cite{Kirchner2020}, where only a small subset of measurements was carefully selected to be sent to the data fusion center. However, greater performance at the fusion center can be observed if a small amount of communication is utilized to coordinate among sensors. Motivated by this observation, this paper seeks to quantify the extent to which inter-sensor coordination, through limited, localized information exchange, can improve fusion performance over the standard setting.

In this paper it is assumed that the objective function the sensors seek to maximize is \emph{monotone submodular}, meaning that more measurements cannot decrease the objective function, and that adding a measurement to an existing set exhibits a ``diminishing returns" property. Monotone submodularity is a common assumption for applications which need to quantify information content in some way \cite{Iyer2021,Krause2008,Badanidiyuru2014}. Therefore, the sensors' global objective is formulated as a submodular maximization problem. Such problems have also been well-studied in the literature, with applications to sensor placement \cite{Krause2008}, outbreak detection in networks \cite{Leskovec2007}, maximizing and inferring influence in a social network \cite{Kempe2003,Gomez-Rodriguez2012}, document summarization \cite{Lin2011}, clustering \cite{mirzasoleiman2016}, assigning satellites to targets \cite{Qu2019}, path planning for multiple robots \cite{Singh2007}, and leader selection and resource allocation in multiagent systems \cite{Clark2011}.

\begin{figure*}[h]
	\centering
	\includegraphics[scale=0.55]{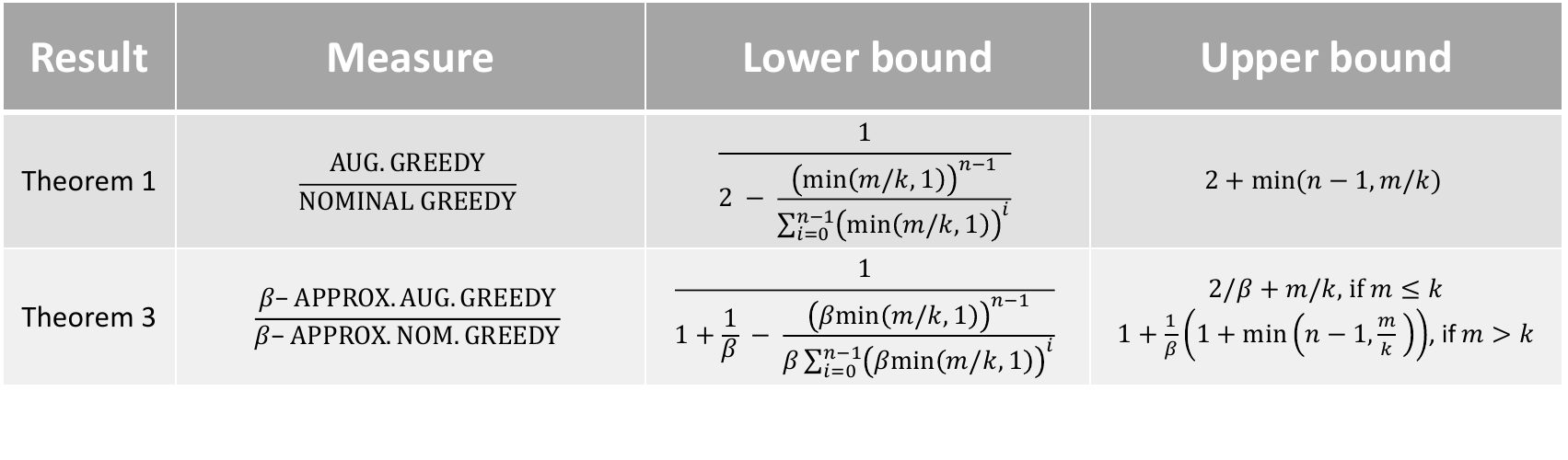}
	\caption{A brief summary of the results from this paper, where a sensor selects up to $k$ measurements as its ``action" and up to $m$ measurements to share. We explore how well an augmented greedy algorithm, which includes measurement passing, performs. Theorem~\ref{thm:eff} provides a range as to how well the augmented greedy algorithm can perform versus the nominal greedy algorithm. Theorem~\ref{thm:approx} describes the scenario where each agent can only approximate the solution to its local problem within a factor of $\beta$.}
	\label{fig:summ}
\end{figure*}

Maximizing a submodular function is an NP-hard problem for many relevant subclasses of submodular functions and constraints \cite{Lovasz1983}. Therefore, much effort has been devoted to finding and improving algorithms which approximate the optimal solution in polynomial time. A key result of this line of research is that algorithms exist which give strong guarantees as to how well the optimal solution can be approximated. One such algorithm is the greedy algorithm, first proposed in the seminal work in \cite{Nemhauser1978a}. Here it was shown that for certain subclasses of problems, (e.g., submodular maximization subject to a uniform matroid constraint) the solution provided by the greedy algorithm must be within a multiplicative factor of $1-1/e \approx 0.63$ of the optimal, and within $1/2$ of the optimal for the more general class of problems \cite{Fisher1978a}. Since then, more sophisticated algorithms have been developed to show that any submodular maximization problem that can be solved efficiently within the $1-1/e$ guarantee \cite{Calinescu2011,Gairing2009}. It has also been shown that progress beyond this level of optimality is not possible using a polynomial-time algorithm, where the indicator step for the time complexity is the number of evaluations of the objective function \cite{Feige1998}.

One line of research has studied distributed methods for solving submodular maximization; see, for instance \cite{Clark2016,robey2021optimal}. The work in \cite{Luo2016DistributedConstraints} addresses a similar problem as this paper, where vehicles are communicating measurements to a centralized entity. The constraints, however, are different from the current setting, and it is shown that an auction-style method of communication among the agents results in a $1-1/e$ guarantee. Of particular interest to this work is a distributed version of the greedy algorithm, where the collective decisions result from a sequential process where partial decisions are made using a defined selection rule \cite{Gharesifard2017}. Each sensor greedily chooses the measurements that provide the most added value relative to those chosen by the previous sensors in the sequence. Like the standard greedy algorithm, it has been shown that this distributed greedy algorithm guarantees the solution is within 1/2 the optimal \cite{Gharesifard2017,Fisher1978a}.

In this setting, recent literature has emerged which explores relaxing the requirement that each sensor has the ability to compute its contribution relative to the decisions of \emph{all} previous sensors. For instance, \cite{Grimsman2019} describes how the 1/2-guarantee decreases as sensors can only access a subset of the measurements chosen by previous sensors. The work in \cite{Grimsman2018} shows how a particular type of communication between the sensors can be employed to recover some of this loss in performance; for example, rather than communicate its own decision to future sensors, a sensor can choose to communicate a more valuable decision of another sensor. Work has also begun to explore how additional knowledge of the structure of the initial measurement distribution can offset this loss \cite{Corah2019,Corah2018}.

This paper addresses the extent to which inter-sensor communications can improve the performance of distributed algorithms. Specifically, we consider a type of submodular maximization problem with $n$ sensors, where each sensor has available a large set of measurements and needs to downselect this large set to a subset of $k$ measurements to be sent to the fusion center. This transmission to the fusion center is overheard by the other sensors (or explicitly communicated to the other sensors). On top of this, we have measurement passing, by which a sensor can forward up to $m$ of its own measurements that subsequent sensors may choose to pass to the fusion center instead (or in combination) to their own measurements. This effectively relaxes the standard constraint in the distributed setting that sensor $i$ can only send measurements from its own local set. As an example, consider a scenario where two flying vehicles are trying to identify the positions of a set of targets. Each vehicle captures many images, but can only send $k$ of them to a central satellite, which uses the sent measurements to estimate the location of the targets. In this scenario, vehicle 1 can use the local communication network to share with vehicle 2 the $k$ images which it sent to the satellite, and can additionally send $m$ more. Vehicle 2 could then send to the satellite any combination of $k$ images from its original set as well as these $m$ new communicated measurements. In the extreme case where vehicle 2 was not able to capture any ``valuable" images by itself, it could still send to the fusion center the $m$ that came from vehicle 1, thus offsetting a potentially poor system performance.

The focus on this paper is to study how the increased communication allowed by measurement passing can improve the value of the resulting set of measurements. Theorem~\ref{thm:eff} addresses this directly by giving bounds on how well a measurement passing policy can perform compared to the distributed greedy algorithm: it shows that there exist problem instances for high $m$ where measurement passing can outperform the greedy algorithm by a multiplicative factor of $n+1$. For smaller $m$, that factor reduces to $2 + m/k$. Theorem~\ref{thm:eff} also shows that any measurement passing algorithm can be outperformed by the greedy algorithm for carefully chosen problem instances, but always by a factor less than 2, and as low as $2-1/n$ for high $m$. 

Theorem~\ref{thm:approx} addresses the practical issue that solving the ``local" problem that each sensor must solve of (i) selecting the $k$ ``best" measurements to send to the fusion center and (ii) selecting the $m$ ``best" measurements to forward to other sensors, can be by themselves intractable problems. This is typically the case for the example scenario described above, when each of the flying vehicles has available a large collection of local measurements. Realistically, the optimal solution to the local problem must be approximated using some computationally-feasible algorithm. Assuming that agents can approximate the solution to the local problem within a factor of $\beta$, Theorem \ref{thm:approx} shows how these local approximations impact the results of Theorem \ref{thm:eff}. Interestingly, these local approximations do not impact the resulting performance guarantees in a significant way. A summary of the theoretical results can be found in Figure~\ref{fig:summ}.

Finally, this paper provides a numerical example to show how measurment passing can improve performance. We show that on average, measurment passing helps the most with few agents (e.g., $n=3$) and many targets (e.g., $10$ targets). However, even with a large number of agents and fewer targets (e.g., $n=10$ and 4 targets), message passing improves performance in the vast majority of the cases (over 99\%).
\section{Model} \label{sec:model}

Consider a set of $n$ sensors, where each sensor $i \in \{1, \dots, n\}$ has a local set of noisy measurements, denoted $S_i$. Each sensor can communicate up to $k$ measurements to a fusion center, where the total $nk$ measurements received will be used to estimate the state of the environment. The collective goal of the sensors is to send the most informative set of measurements to the fusion center so that the estimation is as close as possible to the actual state of the environment.

We model such an environment as a multiagent decision problem where the choice set of each agent $i \in N$ is $(S_i)^k := \{S' \subseteq S: |S'| \le k\}$ for the space of possible measurements $S$ and where the system-level objective is captured by a function $f: 2^S \to \mathbb{R}$. Thus each problem instance can be defined by the tuple $I = (f, S_1, \dots, S_n)$. In this work we assume that the objective function $f$ has the following properties:
\begin{itemize}
	\item \emph{Submodular}: $f(A \cup \{s\}) - f(A) \geq f(B \cup \{s\}) - f(B)$ for all $A \subseteq B \subseteq S$ and $s \in S \setminus B$.
	\item \emph{Monotone}: $f(A) \leq f(B)$ for $A \subseteq B \subseteq S$.
	\item \emph{Normalized}: $f(\emptyset) = 0$.
\end{itemize}
We will refer to such functions simply as \emph{submodular}. In the classical setting, the sensors attempt to solve
\begin{align}
	\label{eq:prob}
	\max_{x_i \in (S_i)^k} f(x_1 \cup \cdots \cup x_n).
\end{align}
This problem is an instance of a submodular maximization problem subject to a partition matroid constraint, a known NP-Hard problem \footnote{Furthermore, it can also be shown that this particular subclass of submodular maximization is itself NP-Hard.}.

\begin{example*}[Flying Vehicles \cite{Kirchner2020}]
	Consider the scenario where the sensors are cameras carried on board $n$ flying vehicles that capture images of ground targets and return their pixel coordinates. Each vehicle $i \in \{1, \dots, n\}$ has access to a large collection of pixel coordinate measurements taken by its own camera, which comprise the local element set $S_i$. However, each vehicle $i$ needs to select a much smaller subset of these measurements (no more than $k$) to send to a satellite for data fusion. The goal of the vehicles is to select the best set of $k$ measurements that each vehicle should send to the satellite so that an optimal estimate $\hat\theta$ of the targets' positions $\theta$ can be recovered by fusing the measurements received from all the vehicles.
	
	To facilitate this goal, one can employ the use of the Fisher Information Matrix $\fim(x)$ for a set of measurements ${x \subseteq S}$, which is defined as follows:
	\begin{equation}
		\fim(x):= Q_0 + \sum_{s \in x} Q_s,
	\end{equation}
	where
	\begin{align}
		Q_0 &:= \frac{\partial \log p(\theta)}{\partial \theta} \cdot \frac{\partial \log p(\theta)}{\partial \theta}^T , \label{eq:q0}\\
		Q_s & := \mathbb{E}_\theta \left[ \left. \frac{\partial \log p(s | \theta)}{\partial \theta} \cdot \frac{\partial \log p(s | \theta)}{\partial \theta}^T \right| \theta \right], \label{eq:qs}
	\end{align}
	and $p(\theta)$ is the a-priori probability density function of $\theta$ and $p(s|\theta)$ is the likelihood of measurement $s \in x$. The positive semidefinite matrices $Q_0$ and $Q_s$ encode the prior information and the informative contribution of measurement $s$, respectively. The FIM is helpful in the current setting, given the Cram\'er-Rao lower bound (CRLB), which states that for an unbiased estimator,
	\begin{equation} \label{eq:crlb}
		\mathbb{E} \left[(\hat\theta(x) - \theta) (\hat\theta(x) - \theta)^T \right] \ge \fim(x)^{-1},
	\end{equation}
	where we use $\ge$ in the sense that if $A \ge B$, then $A-B$ is positive semidefinite. According to \eqref{eq:crlb}, for any optimal estimator that achieves the CRLB, a set $x$ of measurements that ``minimizes" $\fim(x)^{-1}$ also minimizes the error covariance. A scalar metric that is commonly used to measure the information content of a set of measurements is the D-optimality \cite{pazman1986foundations}, which in our context can be defined by
	\begin{equation} \label{eq:fdef}
		f(x) := \log \frac{\det (\fim(x))}{\det(\fim(\emptyset))},
	\end{equation}
	which has been shown to be submodular \cite{Krause2008,Summers2016}.
\end{example*}

\subsection{The Greedy Algorithm}

Although \eqref{eq:prob} is intractable in general, there are simple algorithms that can attain near optimal behavior for this class of submodular optimization problems. One such algorithm, termed the \emph{greedy algorithm} \cite{Fisher1978a}, proceeds according to the following rule: each sensor $i$ sequentially selects its choice $x_i \in (S_i)^k$ by ``greedily'' choosing the action which yields the greatest immediate benefit to the objective $f$, i.e.,
\begin{equation} \label{eq:choice}
	x_i \in \argmax_{\tilde{x} \in (S_i)^k} f\left( x_1 \cup \cdots \cup x_{i-1} \cup \tilde{x} \right).
\end{equation}
While this greedy algorithm can be implemented in a distributed fashion, there is an informational demand on the system, as each sensor $i$ must be aware of the union of the choices of all previous sensors. Further, each sensor must also be able to compute the optimal choice as defined in \eqref{eq:choice}.

While relatively simple, the greedy algorithm is also high-performing as it yields a solution $x = (x_1, \dots x_n)$ which is within 1/2 of the optimal, i.e., $f(x_1 \cup \cdots \cup x_n) \geq (1/2){\rm OPT}$, where $\rm OPT$ is the value of the solution to \eqref{eq:prob} \footnote{It should be noted that this guarantee is also obtained when agents cannot solve \eqref{eq:choice} optimally, but rather implement a local greedy algorithm to choose their actions \cite{Fisher1978a,Goundan2007}. In this work, we begin with the assumption that agents solve \eqref{eq:choice} optimally, and give lower bounds on the guarantee when we relax that assumption in Section \ref{sec:approx}.}.

\subsection{Measurement Sharing} \label{sub:extmodel}

This work explores the notion of \emph{measurement sharing} as an extension of the greedy algorithm, where we relax the constraint that the $k$ measurements sent to the fusion center by sensor $i$ are a subset of $S_i$. Specifically, consider the case where sensor $i$ can share up to $m \geq 0$ of the measurements in $S_i$ to the forthcoming sensors $j > i$, and we denote by $z_i \in (S_i)^m$ those shared measurements. The subsequent sensors $j>i$ can then select their choices from among their original set $S_i$, but also can include some of the shared measurements $z_1 \cup \dots \cup z_{j-1}$ from previous sensors in the sequence. 

In this paper we consider the set of decision-making algorithms of the form $\pi = (\pi_1, \dots, \pi_n)$, where $\pi_i$ is a rule employed by sensor $i$ to select $k \ge 0$ measurements and ``communicate" $m \ge 0$ measurements. Specifically, given the decisions of previous agents $x_1^\pi, \dots, x_{i-1}^\pi$ and measurements shared from previous agents $z_1^\pi, \cdots, z_{i-1}^\pi$, $\pi_i$ specifies both decision and communication of appropriate dimension, i.e.,
\begin{equation} \label{eq:pidef}
	\{x^\pi_i, z^\pi_i\} \in \pi_i^{k,m}(S_i, x^\pi_{1:i-1},
	z^\pi_{1:i-1}),
\end{equation}
subject to the constraint that $x^\pi_i \in (S_i \cup z^\pi_1 \cup \dots \cup z^\pi_{i-1})^k$ and $z_i^\pi \in (S_i)^m$. This constraint ensures that each sensor $i$ can only select elements either from its own set $S_i$ or elements shared by previous sensors $j<i$. We denote $\pi^{k,m}$ to be the set of policies that satisfy \eqref{eq:pidef}, and denote $x^\pi(I)$ to be the resulting decision set for policy $\pi$ on problem instance $I$, abusing notation so that $f(x^\pi(I)) = f(x^\pi_1 \cup \cdots \cup x^\pi_n)$.

Measurement sharing effectively generalizes the nominal greedy algorithm in that it yields solutions which are not in the set $(S_1)^k \times \cdots \times (S_n)^k$. In the same way that the nominal greedy algorithm approximates \eqref{eq:prob}, a measurement passing policy $\pi$ approximates
\begin{equation} \label{eq:prob_mp}
	\max_{\substack{z_i \in (S_i)^m \\ x_i \in (S_i \cup z_1 \cup \cdots \cup z_{i-1})^k}} f(x_1 \cup \cdots \cup x_n)
\end{equation}
Here we give an example policy, which will be analyzed in this paper. 

\begin{definition} [Augmented Greedy Policy] \label{def:ag}
	A policy $\pi$ is an augmented greedy policy if each sensor $i \in \{1, \dots, n\}$ is associated with a selection rule $\pi_i$ of the form
	\begin{subequations} \label{eq:ag_opts}
		\begin{align}
			x_i^\pi & \in \underset{\tilde{x} \subseteq (S_i \cup z_1^\pi \cup \dots \cup z_{i-1}^\pi)^k}{\arg \max} \ f( x^\pi_1 \cup \cdots \cup x^\pi_{i-1} \cup \tilde{x}) \label{eq:choice_ag}\\
			z^\pi_i & = z^k_i \cup z^{m-k}_i, \text{ where} \label{eq:share_ag1} \\
			z^k_i & \in \underset{\tilde{z} \in (S_i)^{\min(m, k)}}{\arg \max}\ f(x^\pi_1 \cup \cdots \cup x^\pi_i \cup \tilde{z}) \label{eq:share_ag2}\\
			z^{m-k}_i & \in \underset{\tilde{z} \in (S_i)^{\max(m - k, 0)}}{\arg \max}\ f(x^\pi_1 \cup \cdots \cup x^\pi_i \cup z^k_i \cup \tilde{z}) \label{eq:share_ag3}
		\end{align}
	\end{subequations}
\end{definition}

\begin{figure}
	\centering
	\begin{subfigure}[b]{0.48\textwidth}
		\centering
		\includegraphics[scale=0.35]{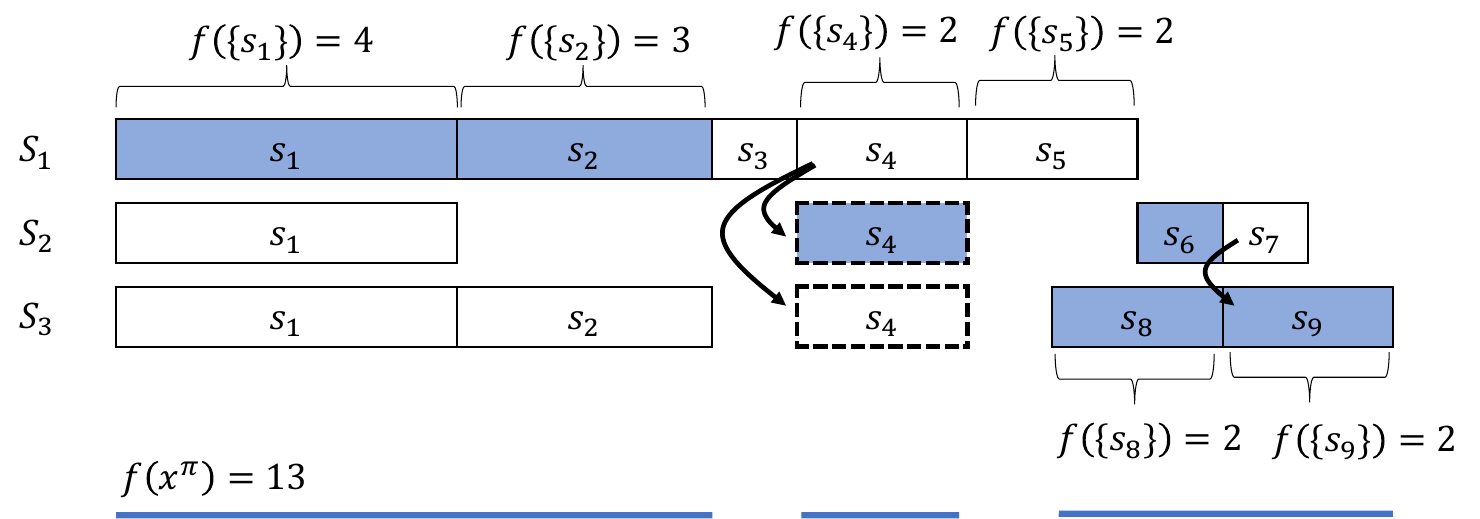}
		\caption{An example problem, where $n=3$, $k=2$, $m=1$. Each box represents an element of $S$, and each row represents $S_i$ for each sensor, i.e., the local measurements to which the sensor has access. The function $f$ is represented by the width of each box, where the width of elements not specifically labeled in the diagram is 1. For $A \subseteq S$, $f(A)$ is the total amount of horizontal space covered by the elements in $A$. For instance, if the agents choose the elements shaded in blue, then $f(x^\pi)$ is represented by the combined lengths of the blue line segments at the bottom of the image. Clearly $f$ is submodular. In this example $\pi$ is an augmented greedy policy. The arrows indicate the measurement passing dictated by $\pi$, for instance $z^\pi_1 = \{s_4\}$. The boxes with the dashed outline indicate that $s_4$ is not in $S_2$ or $S_3$, but is included as part of the sensors' augmented decision set, should they choose to use it.}
		\label{fig:ex}
	\end{subfigure}
	\begin{subfigure}[b]{0.48\textwidth}
		\centering
		\includegraphics[scale=0.5]{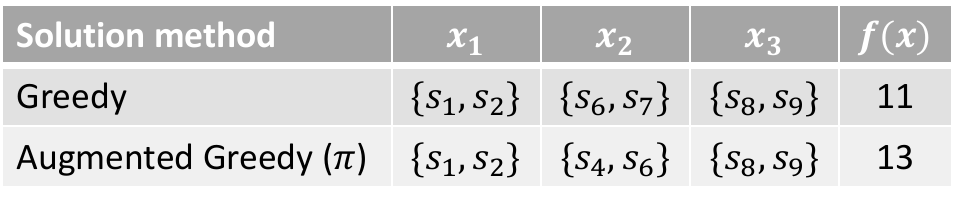}
		\caption{A table representing the performance for 2 different solution methods. First, the greedy algorithm, where sensors choose according to \eqref{eq:choice}, is shown. The second row assumes sensors choose according to an augmented greedy policy.}
		\label{fig:ex_sol}
	\end{subfigure}
	\caption{An example problem illustrating measurement sharing introduced in Section \ref{sub:extmodel}.}
	\label{fig:ill_ex}
\end{figure}

Here each sensor greedily selects the $k$ best measurements for $x_i$ based on what measurements have previously been selected. When $m \le k$, the messages in $z_i$ are the ``next best" $m$ measurements in $S_i$. When $m > k$, $z_i$ is selected in two stages: first, the ``next best" $k$ measurements in $S_i$ are selected, then the ``next best" $m-k$ after that. \footnote{It can be shown that a policy that does not use this two-stage selection rule performs no better than an augmented greedy policy. In particular, as $m \to \infty$, the performance guarantees decrease to the point that message passing gives no advantage.} We note that the rules in \eqref{eq:ag_opts} are not deterministic: the $\argmax$ may be multivalued. Thus there are many augmented greedy policies that satisfy \eqref{eq:ag_opts} in conjunction with some tiebreaking rule. We similarly use the term \emph{nominal greedy policies} in conjunction with the decision rule in \eqref{eq:choice}. See Figure~\ref{fig:ill_ex} for an example problem instance where an augmented greedy policy is used. 

\section{Benchmarking Performance} \label{sec:opt}

In this section, we compare augmented greedy policies against two benchmarks. First, we compare against nominal greedy policies. Since nominal greedy policies are equivalent to the subset of augmented greedy policies where $m=0$, this comparison illustrates how measurement passing can improve system performance. Second, we benchmark against the solution to \eqref{eq:prob}, in order to see how the 1/2 guarantee provided by nominal greedy policies is improved by message passing. We note that in many realistic settings, solving the local optimization problems \eqref{eq:choice} and \eqref{eq:ag_opts} may not be feasible. Here we assume the vehicles can solve these problem precisely in order to highlight the advantage of measurement passing. In Section~\ref{sec:approx} we relax this assumption. 

The first comparison we make between the two classes of policies is direct: we compare the ratio between the two for any given problem instance. We denote $\pi^{k,m}_\ag$ to be the set of augmented greedy policies and $\pi^{k, m}_\nog$ to be the set of nominal greedy policies that satisfy \eqref{eq:choice}. Note that $\pi^{k, 0}_\ag = \pi^{k, 0}_\nog$.

\begin{theorem} \label{thm:eff}
	Consider the measurement selection problem with $n$ sensors. Then for any $k\ge 1, m \ge 0$ the best-case gain in performance and worst-case loss in performance associated with an optimal measurement passing policy within $\pi^{k,m}$ satisfies
	\begin{align}
		&\max_{\pi \in \pi^{k,m}}\max_{I \in \mathcal{I}, \rho \in \pi^{k,m}_{\nog}}\frac{f(x^\pi(I))}{f(x^\rho(I))} \le 2 + \min(m/k, n-1), \label{eq:ub_opt} \\
		&\max_{\pi \in \pi^{k,m}}\min_{I \in \mathcal{I}, \rho \in \pi^{k,m}_{\nog}}\frac{f(x^\pi(I))}{f(x^\rho(I))} \le \frac{1}{2 - \frac{\min((n-1)m/k, 1)}{n-1+\min((n-1)m/k, 1)}}, \label{eq:lb_opt}
	\end{align}
	where $\mathcal{I}$ is the set of all problem instances. When restricting attention to augmented greedy policies, the best-case gain in performance and worst-case loss in performance associated with any $\pi \in \pi^{k,m}_\ag$ satisfies
	\begin{align} 
		&\max_{I \in \mathcal{I}, \rho \in \pi^{k,m}_{\nog}}\frac{f(x^\pi(I))}{f(x^\rho(I))} \ge 2 + \min(m/k, n-1-1/k), \label{eq:ub_ag}\\
		&\min_{I \in \mathcal{I}, \rho \in \pi^{k,m}_{\nog}}\frac{f(x^\pi(I))}{f(x^\rho(I))} \ge \frac{1}{2 - \frac{(\min(m/k, 1)) ^ {n-1}} {\sum_{i=0}^{n-1} (\min( m/k, 1)) ^i}} \label{eq:lb_ag_lb}
	\end{align}
	where the bound in \eqref{eq:ub_ag} becomes an equality of the form \eqref{eq:ub_opt} when $m \le nk - k - 1$ and the bound in \eqref{eq:lb_ag_lb} becomes an equality of the form \eqref{eq:lb_opt} when $m \ge k$.
\end{theorem}

The theorem proof is given in the next subsection. The bounds given in Theorem~\ref{thm:eff} represent a range of possible values for $f(x^\pi(I))/f(x^\rho(I))$ when $\rho \in \pi^{k, m}_\nog$, $\pi \in \pi^{k, m}_\ag$ and for any problem instance $I$. If $m=0$, i.e., $\pi$ is equivalent to a nominal greedy policy, then $2 \ge f(x^\pi(I))/f(x^\rho(I)) \ge 1/2$, since there exist problem instances where there are (at least) two possible outcomes for the greedy algorithm: the solution to \eqref{eq:prob}, and the other a worst-case outcome, which has 1/2 the value of the first outcome. Therefore, one would hope that measurement passing can increase this upper bound above $2$ and lower bound above $1/2$. Theorem~\ref{thm:eff} shows that this is the case.

The second comparison we make between the two classes of policies is indirectly through a benchmark: comparing them each to an optimal solution to \eqref{eq:prob}, to see how measurement sharing increases performance guarantees. When $\pi$ is a nominal greedy policy, as has been stated, the resulting solution is within a factor of $1/2$ of this benchmark. Therefore, we are interested in how the ability to share measurements can increase the $1/2$ bound:
\begin{theorem} \label{thm:lb}
	Consider the measurement selection problem with $n$ sensors. Then for any $k \ge 1$, $m \ge 0$, the worst-case performance guarantee associated with the optimal measurement passing policy within $\pi^{k,m}$ satisfies
	\begin{equation} \label{eq:lb_opt_bench}
		\max _{\pi \in \pi^{k,m}}\min_{I \in \mathcal{I}} \frac{f(x^\pi(I))}{{\rm OPT}(I, k)} \le \frac{1}{2 - \frac{\min((n-1)m/k, 1)}{n-1+\min((n-1)m/k, 1)}},
	\end{equation}
	where ${\rm OPT}(I, k)$ is the value of the solution to \eqref{eq:prob}. When restricting attention to augmented greedy policies, the worst-case performance guarantee associated with any $\pi$ satisfies 
	\begin{equation} \label{eq:lb_ag_bench}
		\min_{I \in \mathcal{I}} \frac{f(x^\pi(I))}{{\rm OPT}(I, k)} \ge \frac{1}{2 - \frac{(\min(m/k, 1)) ^ {n-1}} {\sum_{i=0}^{n-1} (\min( m/k, 1)) ^i}}.
	\end{equation}
	where the bound in \eqref{eq:lb_ag_bench} becomes an equality of the form \eqref{eq:lb_opt_bench} when $m \ge k$.
\end{theorem}
The proof is will be shown below, after some discussion of the above results. It should be clear from Theorem~\ref{thm:lb} that message passing offers a strictly better guarantee against $\rm OPT$. We note that \eqref{eq:lb_opt_bench} is a similar statement to \eqref{eq:lb_opt}, therefore, the discussion of the interpretation of the bound in \eqref{eq:lb_opt} also applies to \eqref{eq:lb_opt_bench}. The same is true for \eqref{eq:lb_ag_bench} and \eqref{eq:lb_ag_lb}. 

Equation \eqref{eq:ub_opt} gives the upper bound on $f(x^\pi(I))/f(x^\rho(I))$ for any $\pi$. As one might expect, this upper bound increases with $m$: the more measurement passing is permitted, the higher the possible performance increase. However, when $m > k(n-1)$, the upper bound remains constant: increasing $m$ above this value no longer increases potential improvement. Equation \eqref{eq:ub_ag} shows that any augmented greedy policy is optimal in this sense when $m \le nk-k-1$. Furthermore, the expressions in \eqref{eq:ub_ag} and \eqref{eq:ub_opt} are always within an additive factor of $1/k$, so any augmented greedy policy is also at least near-optimal in this sense.

 Equation \eqref{eq:lb_opt} shows that for any policy $\pi \in \pi^{k, m}$, one can always carefully construct problem instances where a nominal greedy policy (complete with tie-breaking rule) will perform better. In fact, no policy can guarantee that this worst-case performance loss is higher than the expression in \eqref{eq:lb_opt}. Again, we see that increasing $m$ increases this lower bound, although here one sees no additional increase when $m>k$. Different from the upper bound is that the expression in \eqref{eq:lb_opt} decreases to 1/2 as $n \to \infty$, regardless of the value of $m$. Theorem~\ref{thm:eff} states that any augmented greedy policy is optimal in this sense when $m \ge k$, it is also optimal when $n=2$ or as $n \to \infty$.

\begin{figure}
	\centering
	\begin{subfigure}{0.47\textwidth}
		\centering
		\includegraphics[scale=0.5]{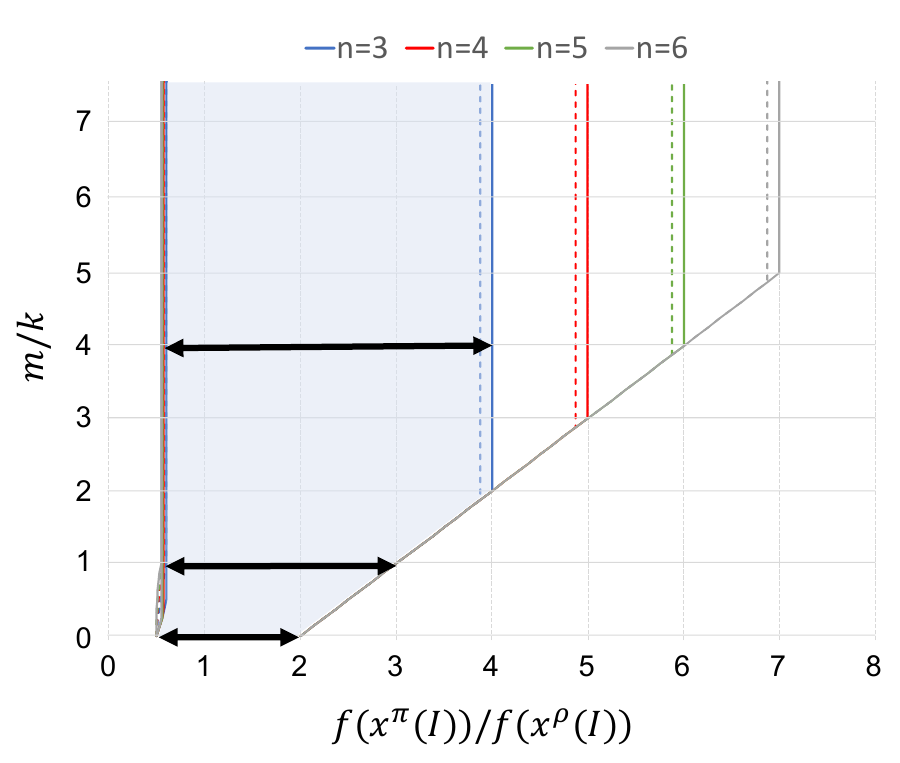}
		\caption{The bounds from Theorem~\ref{thm:eff} for different values of $n$. For any augmented greedy policy $\pi$, the shaded blue region represents all possible values of $f(x^\pi(I))/f(x^\rho(I))$ for the corresponding value of $m/k$ when $n=3$. For instance, the middle of the 3 black solid lines indicates that when $m=k$, $3 \ge f(x^\pi(I))/f(x^\rho(I)) \ge 0.6$.}
		\label{fig:thm1_ub}
	\end{subfigure}
	\begin{subfigure}{0.47\textwidth}
		\centering
		\includegraphics[scale=0.5]{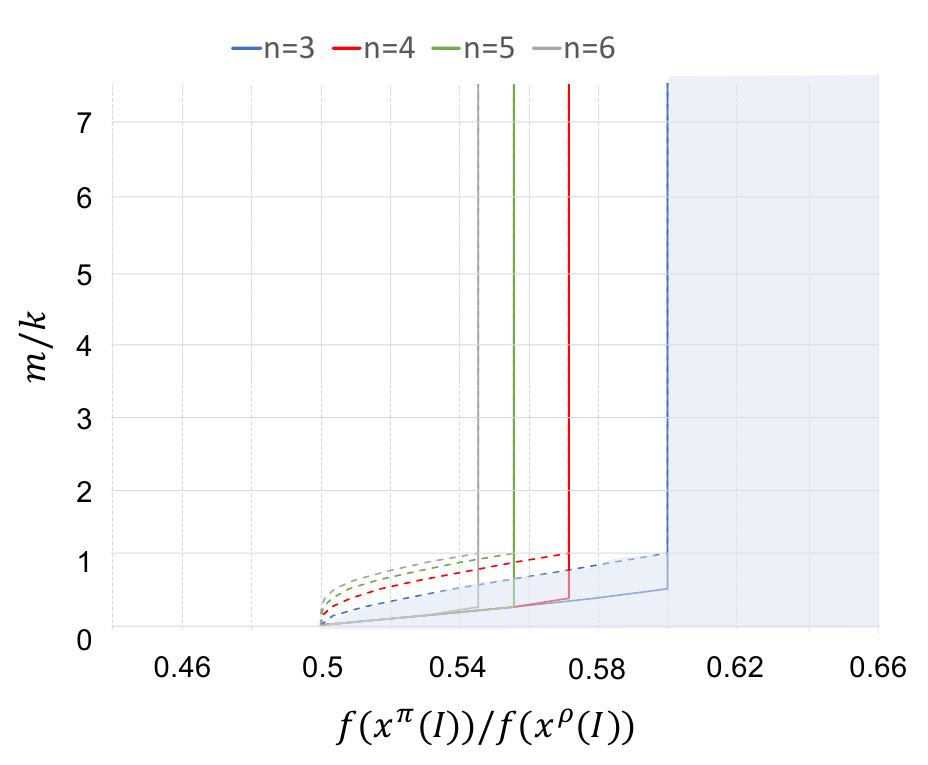}
		\caption{A zoomed-in view of the plot in (a) that shows the lower bounds in \eqref{eq:lb_opt} and \eqref{eq:lb_ag_lb}.}
		\label{fig:thm1_lb}
	\end{subfigure}
	\caption{The results of Theorem~\ref{thm:eff}, illustrated for the case where $k=8$ and for $n=3, 4, 5, 6$. The solid lines indicate the bounds for an optimal message passing policy as given in \eqref{eq:ub_opt} and \eqref{eq:lb_opt} respectively, and the dashed lines indicate the proven lower bounds for any augmented greedy algorithm as given in \eqref{eq:ub_ag} and \eqref{eq:lb_ag_lb}. The plot in (b) is a close-up of the various values for in part (a).}
	\label{fig:eff_plot}
\end{figure}

Figure~\ref{fig:eff_plot} gives an illustration of Theorem~\ref{thm:eff} for the case where $k = 8$ and for $n=3,4,5,6$. The solid colored lines indicate the ``optimal" bounds given in \eqref{eq:ub_opt}--\eqref{eq:lb_opt}, and the dashed lines indicate the bounds for any augmented greedy algorithm as given in \eqref{eq:ub_ag}--\eqref{eq:lb_ag_lb}. For instance, the shaded blue region represents, for $n=3$, the possible values of $f(x^\pi(I))/f(x^\rho(I))$ for any $\pi \in \pi^{k,m}_\ag$, $\pi^{k,m}_\nog$, and $I \in \mathcal{I}$. The lowest solid black line indicates what is described above: that when $m = 0$, the values of the ratio range between 2 and 1/2. The middle solid black line represents the range of values when $m=k$; note that in this region any augmented greedy policy is optimal in that no other policy can provide a higher upper bound or a higher lower bound. Finally, the highest black like represents the range of values when $m = 4k$, where the lower bound is still optimal, but the upper bound is only guaranteed to be near-optimal.

While the performance increase that results from measurement passing is notable, there is some tradeoff with runtime. Note that the nominal greedy policy rule in \eqref{eq:choice} does not prescribe how to solve the local optimization problem, which is intractable in general. A full implementation of the nominal greedy algorithm will require $O\left(n \cdot {\max_i |S_i| \choose k} \right)$ number of calls to $f$. An augmented greedy policy, by comparison, will require $O\left( n \cdot {mn + \max_i |S_i| \choose \max(k, m-k)} \right)$. We address this intractability in Section~\ref{sec:approx}.

Finally, we note that in the examples which serve to prove the bounds in \eqref{eq:lb_opt} and \eqref{eq:ub_ag}, there is some reliance on overlap among the local measurement sets $S_1, \dots, S_n$. Clearly, when $S_i = S_j$ for all $i,j$, then all sensors have access to the same information, thus measurement passing is futile. However, the extent to which this is the case is a topic of future work.

\subsection{Proof for Theorem~\ref{thm:eff}} \label{sub:proof}

We begin with two lemmas that, that, given two policies $\pi, \rho \in \pi^{k, m}$, show marginal contributions for $x_i^\pi$, $z^\pi_i$, and $x_i^\rho$ affects $f(x^\pi)/f(x^\rho)$. We define marginal contribution of $A \subseteq S$ with respect to $B \subseteq S$ according to $f$ as
\begin{equation}
	\Delta(A|B) := f(A \cup B) - f(B).
\end{equation}
We abuse notation so that $\Delta(A, B |C, D) = \Delta(A \cup B | C \cup D)$. Also, denote $x_{a:b} = \cup_{a \le i \le b} x_i$, and likewise for $z_{a:b}$. 

\begin{lemma} \label{lem:ub}
	Assume that policy $\rho \in \pi^{k, m}$ is applied to instance $I \in \mathcal{I}$ and that there exists $\alpha \ge 1$ such that
	\begin{equation} \label{eq:lemub_rho}
		\alpha \Delta(x_i^\rho| x_{1:i-1}^\rho) \ge \max_{\tilde{x} \in (S_i)^k}\Delta(\tilde{x}| x_{1:i-1}^\rho), \ \forall i 
	\end{equation}
	Then for any $\pi \in \pi^{k,m}$,
	\begin{equation} \label{eq:lemub}
		\frac{f(x^\pi(I))}{f(x^\rho(I))} \le 
		\left\{\begin{array}{ll}
			2\alpha + \frac{m}{k}, & \text{ if } m \le k \\ 
			1 + \alpha \left(1 + \min\left(\frac{m}{k}, n-1\right) \right), & \text{ if } m > k.
		\end{array}
		\right.
	\end{equation}
\end{lemma}

\begin{lemma} \label{lem:lb}
	Assume that policy $\pi \in \pi^{k, m}$ is applied to instance $I \in \mathcal{I}$ and that there exist $\alpha_1, \alpha_2 \ge 1$ such that
	\begin{subequations}
		\begin{align}
			& \alpha_1 \Delta(x_i^\pi | x^\pi_{1:i-1}) \ge \max_{\tilde{x} \in (S_i)^k} \Delta(\tilde{x} | x^\pi_{1:i-1}) \label{eq:alpha1} \\
			& \alpha_2 \cdot \max_{\tilde{z} \in (z^\pi_i)^k} \Delta(\tilde{z} | x^\pi_{1:i}) \ge \max_{\tilde{x} \in (S_i)^k} \Delta(\tilde{x} | x^\pi_{1:i-1}). \label{eq:alpha2}
		\end{align}
	\end{subequations}
	Then for any $\rho \in \pi^{k,m}$ such that $x^\rho_i \subseteq S_i$ for all $i$,
	\begin{equation}
		\frac{f(x^\pi(I))}{f(x^\rho(I))} \ge \frac{1}{1 + \alpha_1 - \frac{1}{\sum_{i = 0}^{n-1} \alpha_2^i}}
	\end{equation}
\end{lemma}

The proofs for Lemma~\ref{lem:ub} and Lemma~\ref{lem:lb} are given in Appendix-\ref{app:lemub} and Appendix-\ref{app:lemlb}, respectively. We now prove each statement of the Theorem separately.

\subsubsection{Equation \eqref{eq:ub_opt}}

Here we invoke Lemma~\ref{lem:ub}. Let $\pi \in \pi^{k,m}$ and $\rho \in \pi^{k,m}_\nog$. Using $\rho$, sensors make choices according to \eqref{eq:choice}, therefore for any $I$, let $\alpha=1$. Then both expressions in \eqref{eq:lemub} are equivalent: $f(x^\pi(I)) / f(x^\rho(I)) \le 2 + \min(m/k, n-1)$.

\subsubsection{Equation \eqref{eq:lb_opt}}

\begin{figure}
	\centering
	\begin{subfigure}[b]{0.47\textwidth}
		\centering
		\includegraphics[scale=0.35]{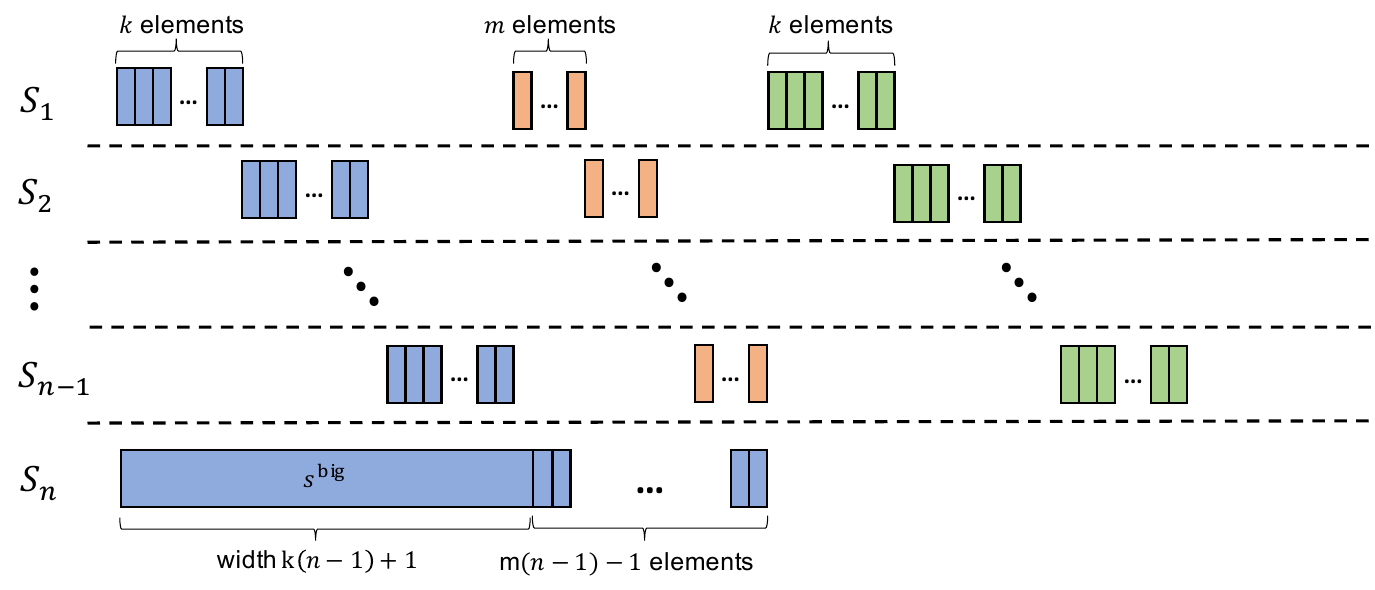}
		\caption{An example for proving \eqref{eq:lb_opt} when $m(n-1) \le k$. The key is that $|S_n| \le k$, and the orange measurements offer no value beyond $S_n$. Each small box has width 1 and $s^{\rm big}$ has width $k(n-1) + 1$. Thus $f(x^\pi(I)) = k(n-1) + m(n-1)$ and $f(x^\rho(I)) = 2k(n-1) + m(n-1)$.}
		\label{fig:ub1}
	\end{subfigure}
	\begin{subfigure}[b]{0.47\textwidth}
		\centering
		\includegraphics[scale=0.35]{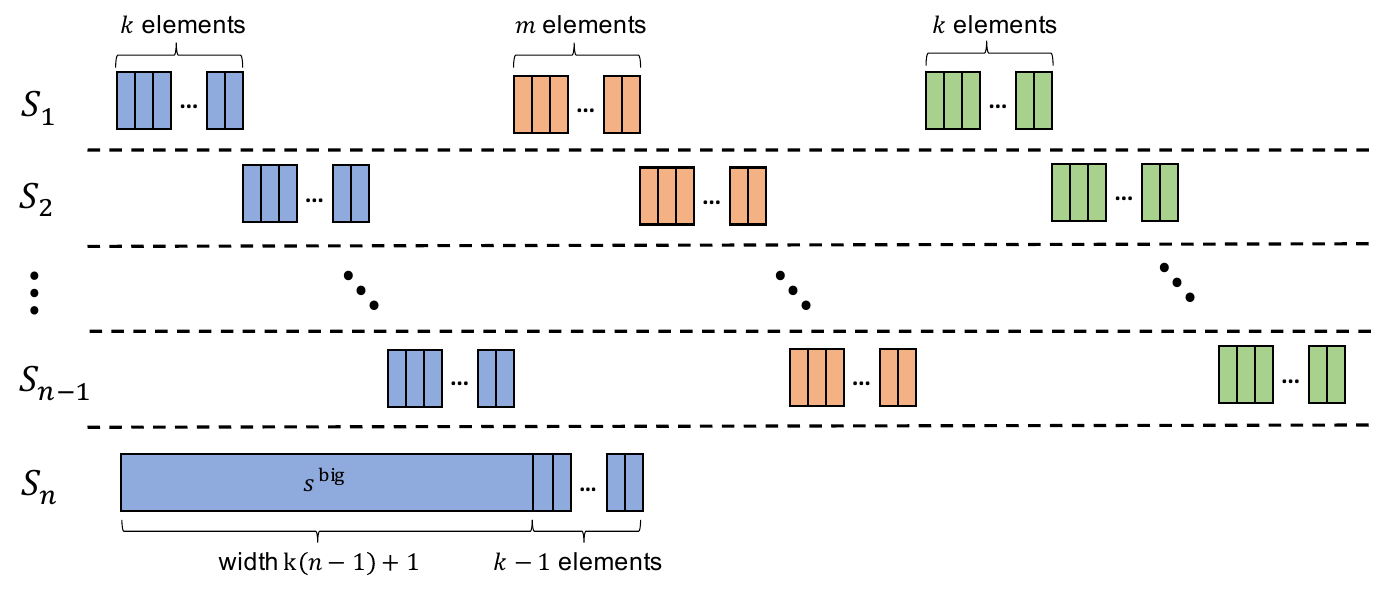}
		\caption{An example for proving \eqref{eq:lb_opt} when $m(n-1) \ge k$. Unlike the example above, here $S_n$ consists of exactly $k$ measurements, thus $f(x^\pi(I)) = k(n-1) + k$ and $f(x^\rho(I)) = 2k(n-1) + k$.}
		\label{fig:ub2}
	\end{subfigure}
	\begin{subfigure}[b]{0.47\textwidth}
		\centering
		\includegraphics[scale=0.35]{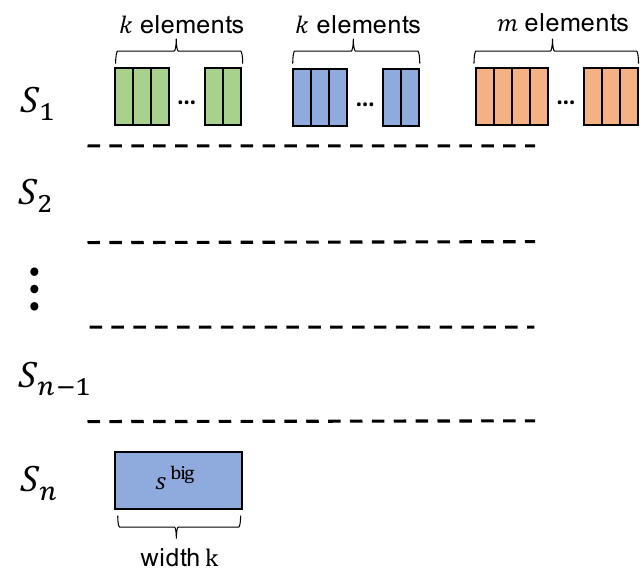}
		\caption{An example for proving \eqref{eq:ub_ag}. Here the greedy algorithm chooses the green measurements which are ``covered" by $s^{\rm big}$, thus $f(x^\rho(I)) = k$ and $f(x^\pi(I)) = 2k + min(m, k(n-1) - 1)$.}
		\label{fig:ub_ag}
	\end{subfigure}
	\caption{Problem instances used in the proof for Theorem~\ref{thm:eff}. This representation is similar to that in Figure \ref{fig:ill_ex}, in that row $i$ represents $S_i$ and $f(A)$ is the amount of horizontal space covered by the boxes in $A \subseteq S$. Blue squares represent measurements chosen by $\pi$, orange are those measurements which are passed using $\pi$, and green are the nominal greedy choices $x^\rho(I)$, when those differ from $x^\pi(I)$. We have omitted the dashed boxes, with the understanding that an orange box in $S_i$ is available to sensors $j>i$.}
	\label{fig:thmeff}
\end{figure}

Fix $\pi \in \pi^{k,m}$ and assume first that $(n-1)m \le k$. Suppose that $f$ and $S_1, \dots, S_n$ are as represented in Figure~\ref{fig:ub1}. Here the format of example is the same as in Figure~\ref{fig:ill_ex}. We assume that all of the small rectangles are of width 1, and that the large rectangle $s^{\rm big}$ is of length $k(n-1) + 1$. Essentially, for sensor $ i \in \{1, \dots, n-1\}$, all measurements in $S_i$ are identical according to $f$, since there is no horizontal overlap among them, and none of these sensors is aware of the measurements in $S_n$.

Assume that $\pi_i$ selects the blue measurements for $x^\pi_i$ and the orange measurements for $z^\pi_i$, $i = 1, \dots, n-1$. Note that $|S_n| = m(n-1) \le k$, so $x_n^\pi = S_n$ is feasible, and an optimal choice regardless of the previous sensors' decisions. This implies that $f(x^\pi(I)) = k(n-1) + m(n-1)$.

On the other hand, consider the decision set of some $\rho \in \pi^{k,m}_\nog$: the rectangles shaded in green for $i<n$, and the blue rectangles for $i=n$. Thus $f(x^\rho(I)) = 2k(n-1) + m(n-1)$, and for this problem instance $I$,
\begin{align}
	\frac{f(x^\pi(I))}{f(x^\rho(I))} = & \frac{k(n-1) + m(n-1)}{2k(n-1) + m(n-1)} \nonumber\\
	= & \frac{1}{2 - \frac{(n-1)m/k}{n-1 + (n-1)m/k}}. \label{eq:lom}
\end{align}

In the case where $m(n-1) \ge k$, consider the example in Figure~\ref{fig:ub2}. Here $S_i$ are the same as in Figure~\ref{fig:ub1}, for $i=1, \dots, n-1$, implying again without that a possible choice for $x^\pi_1, \dots, x^\pi_{n-1}$ are the respective blue measurements. In this case, however, $|S_n| = k$, but note that $f(x^\pi_1 \cup \cdots \cup x^\pi_{n-1} \cup x_n ) = kn$ for any $x_n \in (S_n)^k$, thus $f(x^\pi(I)) = kn$. The green measurements are again a possible greedy policy selection, where different from the blue, showing that $f(x^\rho(I)) = 2k(n-1) + k$. In this case we see that
\begin{align}
	\frac{f(x^\pi(I))}{f(x^\rho(I))} = & \frac{kn}{2k(n-1) + k} = \frac{1}{2 - 1/n}. \label{eq:him}
\end{align}
Equations \eqref{eq:lom} and \eqref{eq:him} establish \eqref{eq:lb_opt} for all cases.

\subsubsection{Equation \eqref{eq:ub_ag}}

We appeal to an example of the same style as in Figure~\ref{fig:ill_ex}, which is illustrated in Figure~\ref{fig:ub_ag}. Here $S_1$ is the union of a set of $k$ green measurements, a set of $k$ blue measurements, and a set of $m$ orange measurements. The sets $S_2, \dots, S_{n-1}$ are all empty, and $S_n = \{s^{\rm big}\}$. All measurements have value 1, except $s^{\rm big}$, which has value $k$. The measurement $s^{\rm big}$ ``covers" the set of green measurements, i.e., if $A$ is the set of green measurements, then $f(B \cup \{s^{\rm big}\}) = k$ for any $B \subseteq A$.

Assume that using a nominal greedy policy $\rho$, sensor 1 selects the $k$ green measurements. Then $f(x^\rho(I)) = k$. However, there exists an augmented greedy policy $\pi$ such that sensor 1 selects the $k$ blue measurements as $x^\pi_1$ and the $m$ orange measurements as $z^\pi_1$. Since the remaining sensors have no other alternatives, $\min\{m, k(n-1)-1\}$ orange measurements are chosen for $x^\pi_2, \dots, x^\pi_{n}$. This implies that $f(x^\pi(I)) = 2k +\min\{m, k(n-1)-1\}$, and that for this problem instance $I$
\begin{equation}
	\frac{f(x^\pi(I))}{f(x^\rho(I))} = 2 + \min\{m/k, n-1-1/k\}.
\end{equation}

\subsubsection{Equation \eqref{eq:lb_ag_lb}}

We invoke Lemma~\ref{lem:lb} by finding acceptable values of $\alpha_1, \alpha_2$ which hold for any $I$. For any $\pi \in \pi^{k,m}_\ag$, \eqref{eq:choice_ag} implies that $\alpha_1 = 1$ is a valid parameter choice. When $m \geq k$, $\alpha_2 = 1$, since $\max_{\tilde{z} \in (z_i^\pi)^k} \Delta(\tilde{z}|x_{1:i}^\pi) = z^k_i$ from \eqref{eq:share_ag2}. When $m < k$, the following holds:
\begin{subequations}
	\begin{align}
		\max_{\tilde{z} \in (z_i^\pi)^k} \Delta(\tilde{z}|x_{1:i}^\pi) = & \max_{\tilde{z} \in (S_i)^m} \Delta(\tilde{z}|x_{1:i}^\pi) \label{eq:lb_ag1} \\
		\ge & (m/k) \cdot \max_{\tilde{x} \in (S_i)^k} \Delta(\tilde{x}|x_{1:i}^\pi). \label{eq:lb_ag2}
	\end{align}
\end{subequations}
Therefore, $\alpha_2 = 1/\min(m/k, 1)$ is an acceptable value. Since $\rho \in \pi^{k,m}_\nog$ satisfies that $x^\rho_i \subseteq S_i$, one can use these values for $\alpha_1, \alpha_2$ (combined with some algebraic manipulation), so that Lemma~\ref{lem:lb} implies \eqref{eq:lb_ag_lb}.

\hfill $\qed$

\subsection{Proof for Theorem~\ref{thm:lb}}

This result follows from Theorem~\ref{thm:eff}: \eqref{eq:lb_opt_bench} since the defining example in Figures~\ref{fig:ub1} and \ref{fig:ub2} can be altered so that the green measurements are the solution to \eqref{eq:prob}, and \eqref{eq:lb_ag_bench} since a policy $\rho \in \pi^{k, m}$ which finds ${\rm OPT}(I, k)$ satisfies the requirement to apply Lemma~\ref{lem:lb}. 

\hfill $\qed$
\section{Suboptimal Selections} \label{sec:approx}

To implement an augmented greedy policy, each agent is required to solve the optimizations in \eqref{eq:choice_ag} and \eqref{eq:share_ag2}, both of which are NP-Hard, in terms of $k$, assuming large $|S_i|$. This means that, when $k$ and $|S_i|$ are large, executing an augmented greedy policy may become computationally infeasible---an observation which is well-known for the nominal greedy algorithm \cite{Goundan2007}. Such scenarios are typical for the the motivating example in Section~\ref{sec:model}, in which the flying vehicles may need to select a large number of images from a much larger set of total images taken. 

Real-world implementations of the augmented greedy algorithm must thus approximate \eqref{eq:choice_ag}--\eqref{eq:share_ag3}. We devote this section to understanding how approximating the solution to such optimizations affects the measurement sharing. The key observation from the results in that follow is that while this approximation increases the range of possible values for $f(x^\pi(I)) / f(x^\rho(I))$ (as one might expect), the lower bound decreases (roughly) linearly as a factor of the approximation error. The idea of using approximate maximization for the nominal greedy algorithm has been used previously in the literature (see, for instance \cite{Goundan2007,lehmann2006combinatorial}) for analyzing approximations to the nominal greedy algorithm, which model and results we extend here for augmented greedy policies.

\begin{definition}[$\beta$-Greedy Policy]
    A policy $\bar{\pi} \in \bar{\pi}^{k,m}$ is a $\beta$-greedy policy for some $\beta \ge 1$ if each sensor $i \in \{1, \dots, n\}$ is associated with a selection rule $\bar{\pi}_i$ of the form
    \begin{equation} \label{eq:choice_ang}
        \Delta(x^{\bar{\pi}}_i | x_{1:i-1}^{\bar{\pi}}) \ge \beta \cdot \max_{\tilde{x} \in (S_i)^k} \Delta(\tilde{x} | x_{1:i-1}^{\bar{\pi}}).
    \end{equation}
    Note that when $\beta=1$, the nominal greedy policy defined by \eqref{eq:choice} is recovered.
\end{definition}

An analogous approximation can be defined for the augmented greedy algorithm:

\begin{definition}[($\beta_k,\beta_m$)-Augmented Greedy Policy] \label{def:appx}
    A policy $\bar{\pi} \in \bar{\pi}^{k,m}$ is a $(\beta_k, \beta_m)$-greedy policy for some $\beta_k,\beta_m \ge 1$ if each sensor $i \in \{1, \dots, n\}$ is associated with a selection rule $\bar{\pi}_i$ of the form
    \begin{subequations}
    \begin{align}
        & \Delta(x_i^{\bar{\pi}} | x_{1:i-1}^{\bar{\pi}}) \ge \beta_k \cdot \underset{\tilde{x} \in (S_i \cup z_1^{\bar{\pi}} \cup \dots \cup z_{i-1}^{\bar{\pi}})^k}{\max} \  \Delta(\tilde{x} | x^{\bar{\pi}}_{1:i-1}), \label{eq:choice_apx}\\
        & z^{\bar{\pi}}_i = z^k_i \cup z^{m-k}_i, \text{ where} \label{eq:share_apx1} \\
        & \Delta(z^k_i | x^{\bar{\pi}}_{1:i}) \ge \beta_m \cdot \underset{\tilde{z} \in (S_i)^{\min(m, k)}}{\max}\ \Delta(\tilde{z} | x^{\bar{\pi}}_{1:i}), \label{eq:share_apx2}\\
        & \Delta(z_i^{m-k} | x^{\bar{\pi}}_{1:i}, z^k_i) \ge \beta_m \cdot \underset{\tilde{z} \in (S_i)^{\max(m-k,0)}}{\max}\ \Delta(\tilde{z} | x^{\bar{\pi}}_{1:i}, z^k_i) \label{eq:share_apx3}
    \end{align}
    \end{subequations}
    In essence, a ($\beta_k,\beta_m$)-augmented greedy policy is a policy which approximates an augmented greedy policy by finding a solution to \eqref{eq:choice_ag} and $\eqref{eq:share_ag2}$ within a factor of $\beta_k$ and $\beta_m$, respectively, of the optimal. When $\beta_k=\beta_m=1$, the original augmented greedy policy is recovered.  
\end{definition}

Alternatively stated. $\beta_m$ is related to how well one can approximate the maximum on $S_i$ and $\beta_k$ is related to how well one can approximate the maximum on $S_i \cup z_1 \cup \cdots \cup z_{i-1}$.

\begin{theorem} \label{thm:approx}
	Consider the measurement selection problem with $n$ sensors. Then for any $(\beta_k, \beta_m)$-augmented greedy policy $\bar{\pi}$, any $\beta_k$-greedy policy ${\bar{\rho}}$, and any problem instance $I$,
	\begin{align} 
		&\frac{f(x^{\bar{\pi}}( I))}{f(x^{\bar{\rho}}(I))} \le 
		\left\{\begin{array}{ll}
			\frac{2}{\beta_k} + \frac{m}{k}, & \text{ if } m \le k \\ 
			1 + \frac{1}{\beta_k}\left(1 + \min\left(\frac{m}{k}, n-1\right) \right), & \text{ if } m > k,
		\end{array} \right. . \label{eq:apxub} \\
		&\frac{f(x^{\bar{\pi}}( I))}{f(x^{\bar{\rho}}(I))} \ge \frac{1}{1 + \frac{1}{\beta_k} - \frac{(\beta_m\min(m/k, 1))^{n-1}}{\sum_{i=0}^{n-1}(\beta_m\min(m/k,1))^i}}. \label{eq:apxlb}
	\end{align}
\end{theorem}

Observe that when $\beta_k = \beta_m = 1$, the results are equivalent to \eqref{eq:ub_opt} and \eqref{eq:lb_ag_lb} in Theorem~\ref{thm:eff}. Here we forgo analogous results to \eqref{eq:lb_opt} and \eqref{eq:ub_ag}, since the emphasis of this theorem is that while the approximations to the local optimization problems increase the range of possible values for $f(x^{\bar{\pi}}( I)) / f(x^{\bar{\rho}}(I))$, this range is still desirable. For instance, if  $\beta_k=\beta_m=1/2$, regardless of the number of sensors, Theorem~\ref{thm:approx} shows that $f(x^{\bar{\pi}}( I)) / f(x^{\bar{\rho}}(I)) \ge 1/3$, as compared to the $1/2$ bound from Theorem~\ref{thm:eff}, i.e., the potential loss in performance decreases only a moderate amount. On the other hand, $f(x^{\bar{\pi}}( I)) / f(x^{\bar{\rho}}(I)) \le 2n+1$, an increase over the $n+1$ bound from Theorem~\ref{thm:eff}. In words, measurement passing can offer a potentially larger benefit without a much higher risk.

An example of a ($\beta_k, \beta_m$)-augmented greedy policy $\bar{\pi}$ is where agent $i$ chooses $x^{\bar{\pi}}_i = \{s_1, \dots, s_k\}$ by sequentially selecting element $s_l$ with the following method:
\begin{equation} \label{eq:seq}
	s_l \in \argmax_{\tilde{s} \in S_i \cup z_1^{\bar{\pi}} \cup \cdots z_{l-1}^{\bar{\pi}}} \Delta(\{\tilde{s}\},s_{1:l-1} | x_{1:i-1}^{\bar{\pi}}).
\end{equation}
This is yet another variation of greedy algorithm, this time for choosing the $k$ elements of $x^{\bar{\pi}}_i$. The guarantees for this algorithm are such that $\beta_k = 1 - (1-1/k)^k$ \cite{Nemhauser1978a}. Using a similar method to choose $z_i^{\bar{\pi}}$ yields $\beta_m = 1 - (1-1/m)^m$, so that both $\beta_k$ and $\beta_m$ are greater than $1-1/e \approx 0.63$.  The $(1-1/e, 1-1/e)$-augmented greedy algorithm can now be implemented using $O(mn^2 + \sum_i |S_i|)$ calls to $f$, and the $(1 - 1/e)$-greedy algorithm can be implemented using $O(\sum_i |S_i|)$ calls to $f$.

\begin{figure}
	\centering
	\includegraphics[scale=0.3]{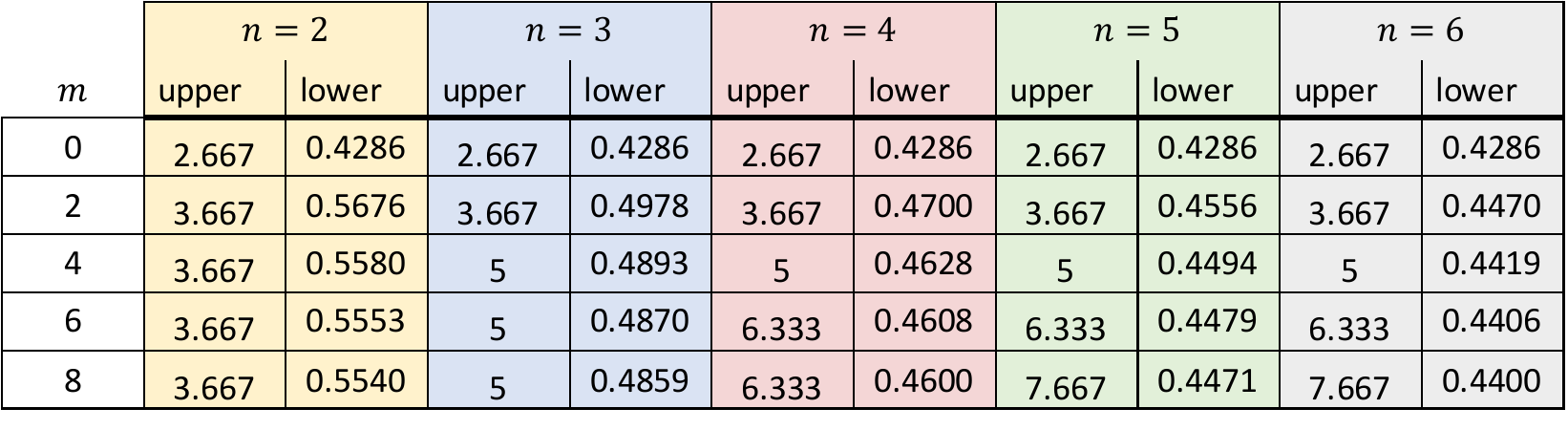}
	\caption{Some examples that showcase the results of Theorem~\ref{thm:approx} for $k=2$. It is assumed that each algorithm is implemented with the sequential greedy rule in \eqref{eq:seq}, i.e., $\beta_k = 1 - (1-1/k)^k$ and $\beta_m = 1 - (1-1/m)^m$.}
	\label{fig:approx_ex}
\end{figure}

Figure~\ref{fig:approx_ex} illustrates the upper bound shown in \eqref{eq:apxub} and the lower bound shown in \eqref{eq:apxlb} for various values of $n$ and $m$ when $k=2$. Here we assume that \eqref{eq:seq} is used to implement both algorithms, i.e., $\beta_k = 1 -(1 - 1/k)^k$ and $\beta_m = 1 - (1 - 1/m)^m$. Since $\beta_m$ decreases as $m$ increases, the lower bound decreases when $m > k$. The upper bound, however, continues to increase in a similar manner to that of Theorem~\ref{thm:eff}.

We now give the proof for Theorem~\ref{thm:approx}:

\begin{proof}
    We first show \eqref{eq:apxlb} by invoking Lemma~\ref{lem:lb}: it  suffices to show valid values for $\alpha_1, \alpha_2$  in order to show the lower bound in \eqref{eq:apxlb}. An immediate consequence of \eqref{eq:choice_apx} is that $\alpha_1 = 1/\beta_k$ holds for all $I$. Likewise, one can use a similar argument to \eqref{eq:lb_ag1}--\eqref{eq:lb_ag2} to show that $\alpha_3 = 1 / (\beta_m \min (m/k, 1))$ holds for all $I$. Then by Lemma~\ref{lem:lb} (since again ${\bar{\rho}}$ is such that $x^{\bar{\rho}}_i \subseteq S_i$),
    \begin{align*}
        \frac{f(x^{\bar{\pi}}(I))}{f(x^{\bar{\rho}}(I))} \ge & \frac{1}{1 + \frac{1}{\beta_k} - \frac{1}{\sum_{i=0}^{n-1}(\beta_m\min(m/k, 1))^i}} \\
        =& \frac{1}{1 + \frac{1}{\beta_k} - \frac{(\beta_m\min(m/k, 1))^{n-1}}{\sum_{i=0}^{n-1}(\beta_m\min(m/k, 1))^i}}.
    \end{align*}

	Equation \eqref{eq:apxub} can be shown similarly using Lemma~\ref{lem:ub}, where $\bar{\pi}$ is the $(\beta_k, \beta_m)$-augmented greedy algorithm, ${\bar{\rho}}$ is the $\beta_k$-greedy algorithm, and, by \eqref{eq:choice_ang}, $\alpha=1/\beta_k$ for all $I$.
\end{proof}
\section{Numerical Example} \label{sec:sim}

\begin{figure}
	\centering
	\includegraphics[scale=0.4]{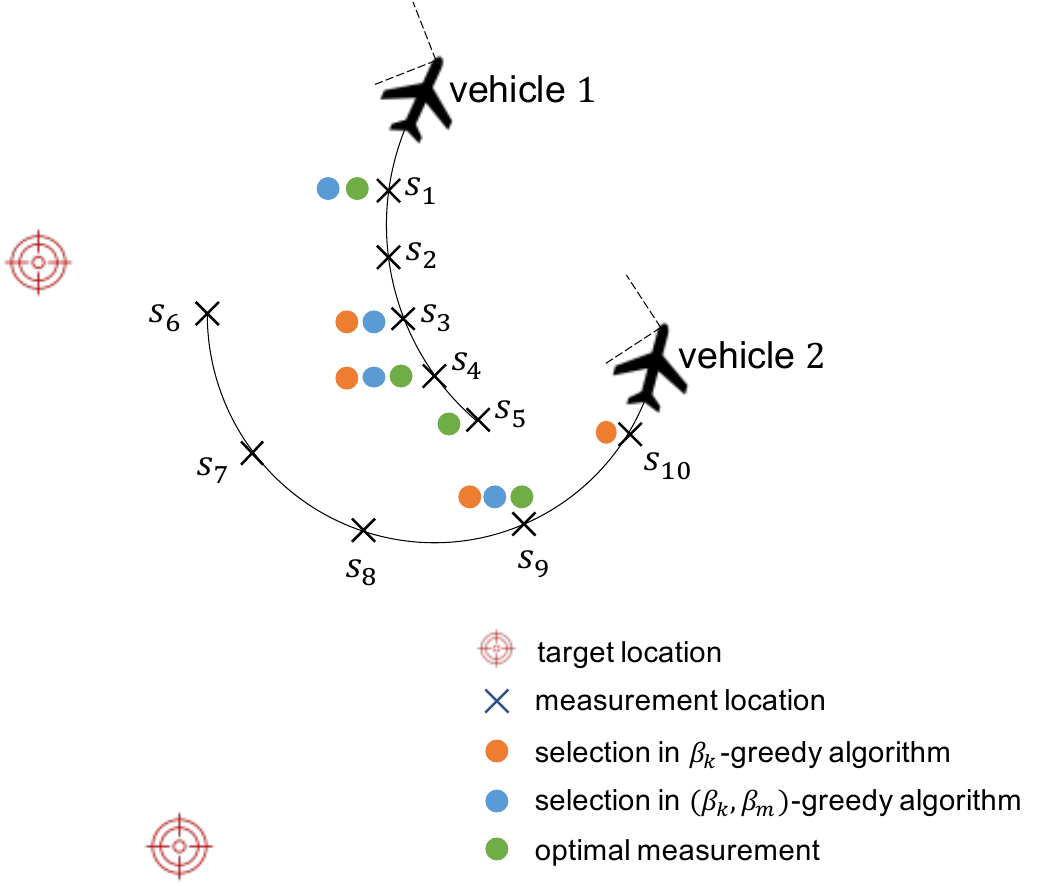}
	\caption{An instance of the flying vehicles problem where $n=2$, $k=2$, $m=1$, and $|S_1|=|S_2|=5$.  The black solid lines indicate each vehicle's path, and the black dashed lines indicate each vehicles field of view. The black $\times$'s are where measurements of the red targets are taken; of course, both targets are not always in view of both vehicles. The $\beta_k$-greedy (orange dots) and $(\beta_k, \beta_m)$-augmented (blue dots) greedy policies are used in conjunction with the sequential selection rule in \eqref{eq:seq}. The green dots indicate measurements that are optimal in the sense that they are a solution to \eqref{eq:prob_mp}.}
	\label{fig:flying_vehicles}
\end{figure}

In this section, we present results for instances of the flying vehicles problem in Section \ref{sec:model}, where $n=2$ flying vehicles move on a curved path, each carrying a side-looking camera with a 90$^\circ$ field of view, a $50$ pixel focal length, and measurement noise in the image plane with standard deviation $\sigma=1$ pixel. There are two stationary ground targets whose 2-D location is to be estimated using the images collected by the flying vehicles. A large number of instances were created with the two targets uniformly randomly placed in the square $[-100,100]\times[-100,100]$. The start position, direction, and turn rate of the each flying vehicle's path were also chosen uniformly randomly. Each flying vehicle moves at a constant forward speed and collects 100 independent measurements uniformly along its path. Details of how to construct the corresponding matrices $Q_0$ and $Q_s$ in \eqref{eq:q0} and \eqref{eq:qs} that quantify the information gain of camera measurements are found in \cite{Kirchner2020}. See Figure \ref{fig:flying_vehicles} for an example.

\begin{figure}
	\centering
	\includegraphics[scale=0.4]{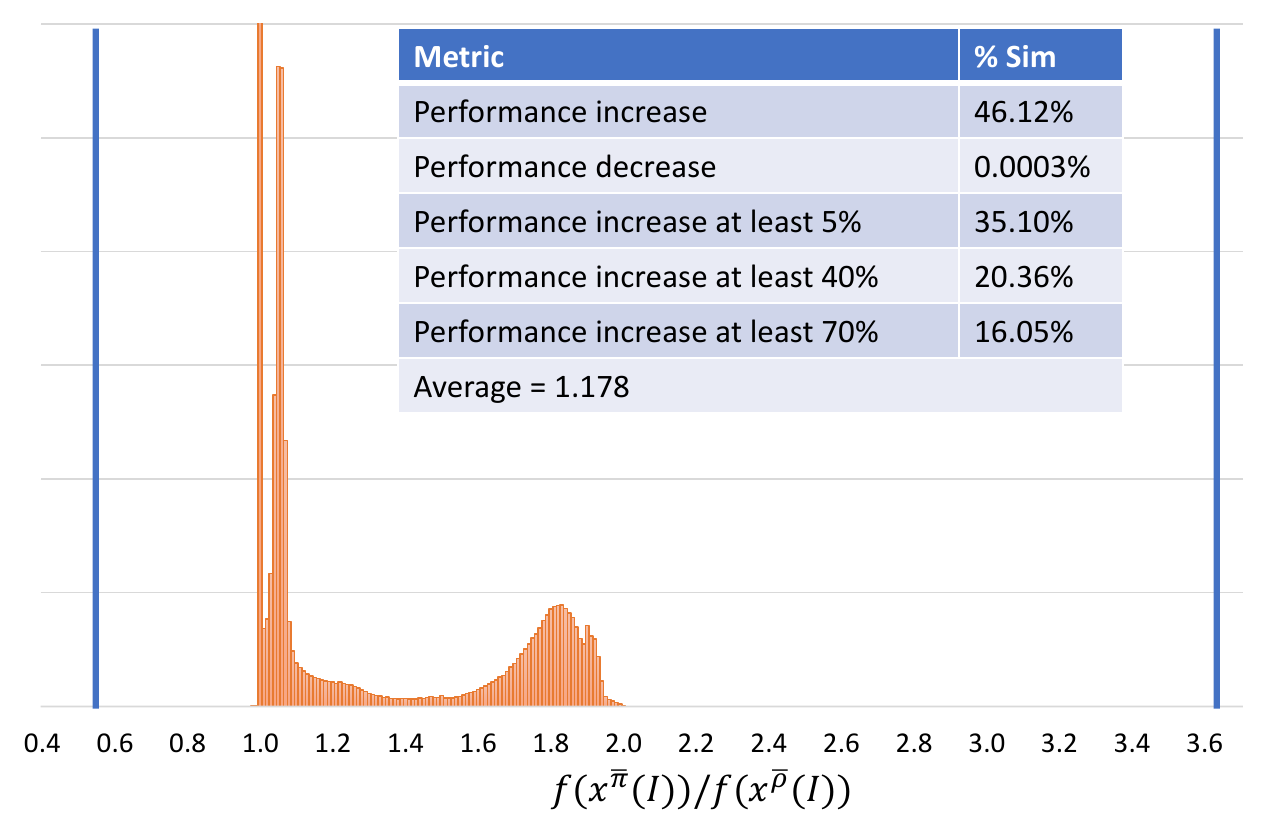}
	\caption{Histogram of the relative performance for $10^6$ random simulations of the 2 vehicle measurement selection problem. A large number of samples fall at 1, which is interpreted as the two policies performing the same. Instances where the value is greater than one indicate the $(\beta^k, \beta^m)$-augmented greedy performed better than the $\beta^k$-greedy in the simulated scenario. The height of the bin corresponding to the ratio of 1 has been cropped. The solid blue bars represent the range of values for $f(x^{\bar{\pi}}(I)) / f(x^{\bar{\rho}}(I))$ as determined by Theorem~\ref{thm:approx} and in Figure~\ref{fig:approx_ex}.}
	\label{fig:opt_real}
\end{figure}

Figure \ref{fig:opt_real} summarizes the results in terms of the ratio between the performance of a $(1-(1-1/k)^k, 1-(1-1/m)^m)$-augmented greedy policy and a $1-(1-1/k)^k$-greedy policy, where $k=m=n=2$ and $|S_1|=|S_2|=100$. Because the number of measurements is very large, the optimizations in \eqref{eq:choice_ag}, \eqref{eq:share_ag2}, and \eqref{eq:choice} are all approximated by the sequential algorithm in \eqref{eq:seq}, where all ties are broken by the index of the measurement.

\begin{figure}
	\centering
	\includegraphics[scale=0.6]{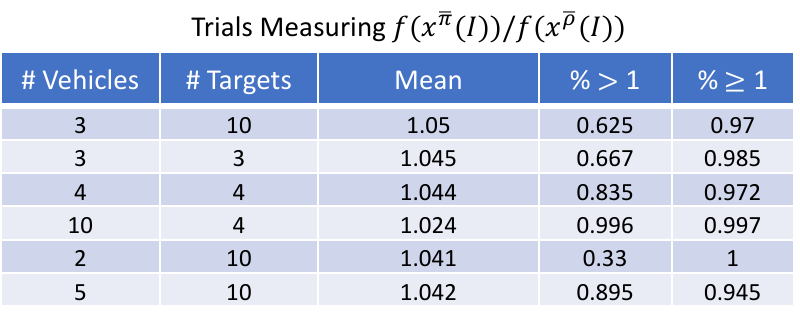}
	\caption{Results from many different types of trials comparing nominal greedy to augmented greedy. Each row represents a different set of numbers of agents and targets, and there were $10^5$ trials run for each row. One can observe that measurement passing helps on average more when there are fewer agents, however, many agents ensure that measurement passing helps more frequently.}
	\label{fig:trials+}
\end{figure}

Additional trials were run for different combinations of the number of flying vehicles and targets. Each combination was repeated $10^5$ times with random position and paths. The results are summarized in Figure~\ref{fig:trials+}. The setup is the same as above, with the exception that, for the purposes of computation the flying vehicles gather 20 measurements over the course of the simulation rather than 100. One can see that when there are more sensors, the mean benefit of measurement passing decreases, since the measurement passing is most useful when a large percentage of flying vehicles cannot observe any targets. This is more likely to happen with fewer flying vehicles. However, one can also observe from the column labeled ``\% $> 1$" that measurement passing is more likely to give some benefit (although smaller) when there are more vehicles, simply because more measurements are being passed, increasing the likelihood that some vehicle will use another's measurement. Likewise, it can be observed from the last column that systems with many sensors are less likely to see a decrease in performance due to measurement passing. 

While measurement selection is based on the submodular function~\eqref{eq:fdef}, it is worth noting that the actual improvement in target estimation, such as that described by the commonly used D-optimality estimation criterion of $\det \left( Q_0 + \sum_{s\in S} Q_s \right)$ (without the log function), can be much greater than that shown in Figure~\ref{fig:opt_real}.

\section{Conclusion}

In this paper we have shown how measurement passing affects the performance guarantees of a group of sensors sequentially selecting measurements. We also showed that the augmented greedy policy gives optimal performance guarantees for some problem instances and near-optimal guarantees for others. Using such a policy, this paper explored how much performance could increase or decrease for any problem instance, and showed how both of these results are affected by inability of each sensor to solve its own local optimization problem.

Future work will continue to explore measurement passing, first by asking which is more important: relaying a sensor's decision or measurement passing? Some preliminary exploration has shown that the sensor's decision is more impactful, but there is not a formal proof of this. Another direction is to apply these results to situations where sensors can only observe the measurements of a subset of previous sensors, and again ask questions about what measurements should be shared and selected.

\bibliographystyle{IEEEtran}
\bibliography{references,util/refs,util/IntroRefs}

\appendix

\subsection{Proof for Lemma \ref{lem:ub}} \label{app:lemub}

Begin with the following:
\begin{subequations}
	\begin{align}
		f&(x^\pi) \leq f(x^\rho) + \Delta(x^\pi | x^\rho) \label{eq:posub2} \\
		\leq & f(x^\rho) + \Delta(z_{1:n-1}^\pi \cap x^\pi | x^\rho) + \Delta(x^\pi \setminus z_{1:n-1}^\pi | x^\rho) \label{eq:posub3} \\
		= & f(x^\rho) + \Delta(z_{1:n-1}^\pi \cap x^\pi | x^\rho) \nonumber \\
		& +\sum_i \Delta(x^\pi_i \setminus z_{1:n-1}^\pi | x^\rho, x^\pi_{1:i-1} \setminus z_{1:n-1}^\pi) \label{eq:posub16} \\
		\le & f(x^\rho) + \Delta(z_{1:n-1}^\pi \cap x^\pi | x^\rho) \nonumber \\
		& +\sum_i \Delta(x^\pi_i \setminus z_{1:n-1}^\pi | x^\rho_{1:i-1}) \label{eq:posub17} \\
		\le & f(x^\rho) + \Delta(z_{1:n-1}^\pi \cap x^\pi | x^\rho) + \sum_i \alpha\Delta(x^\rho_i | x^\rho_{1:i-1}) \label{eq:posub18} \\
		= & (1 + \alpha)f(x^\rho) + \Delta(z_{1:n-1}^\pi \cap x^\pi | x^\rho) \label{eq:posub15}
	\end{align}
\end{subequations}
where \eqref{eq:posub2}, \eqref{eq:posub3}, \eqref{eq:posub17} are true by submodularity of $f$, and \eqref{eq:posub18} is true by \eqref{eq:lemub_rho}.

We denote $\tilde{z}^\pi_i$ to mean $z^\pi_i \cap x^\pi$, and suppose that there exists $\beta$ such that $\Delta(\tilde{z}^\pi_i | x^\rho_{1:i}) \le \beta\Delta(x^\rho_i | x^\rho_{1:i-1})$. Then the second term in \eqref{eq:posub15} can be upper bounded by the following:
\begin{subequations}
	\begin{align}
		\Delta(\tilde{z}_{1:n-1} | x^\rho) \le & \sum_{i=1}^{n-1} \Delta(\tilde{z}^\pi_i | \tilde{z}^\pi_{1:i-1}, x^\rho) \label{eq:posub10}\\
		\le & \sum_{i=1}^{n-1} \Delta(\tilde{z}^\pi_i | x^\rho_{1:i}) \label{eq:posub11} \\
		\le & \sum_{i=1}^{n-1} \beta \Delta(x^\rho_i | x^\rho_{1:i}) \label{eq:posub19} \\
		= & \beta f(x^\rho_{1:n-1}) \le \beta f(x^\rho) \label{eq:posub21},
	\end{align}
\end{subequations}
where \eqref{eq:posub11} and \eqref{eq:posub21} are true by the submodularity of $f$. Substituting this upper bound back into \eqref{eq:posub15}, we see that
\begin{equation*}
	f(x^\pi) \le (1 + \alpha + \beta) f(x^\rho).
\end{equation*}

We now show that such a $\beta$ exists and define it for two cases: when $m \le k$ and when $m \ge k$. First suppose that $m \le k$. Denote $x^{k-m}_i \in \argmax_{\tilde{x} \in (x_i^\rho)^{k-m}} \Delta(\tilde{x} | x^\rho_{1:i-1})$. Then
\begin{subequations}
	\begin{align}
		\Delta( & x^\rho_i | x^\rho_{1:i-1}) \ge (1/\alpha) \Delta(\tilde{z}^\pi_i \cup x^{k-m}_i | x^\rho_{1:i-1}) \label{eq:posub22}\\
		= & (1/\alpha)\Delta(\tilde{z}^\pi_i | x^\rho_{1:i-1}, x_i^{k-m}) + (1/\alpha)\Delta(x^{k-m}_i | x^\rho_{1:i-1}) \label{eq:posub23} \\
		\ge & (1/\alpha)\Delta(\tilde{z}^\pi_i | x^\rho_{1:i}) + (1/\alpha)\Delta(x^{k-m}_i | x^\rho_{1:i-1}) \implies \label{eq:posub24}
	\end{align}
	\begin{align}
		\Delta(& \tilde{z}^\pi_i | x^\rho_{1:i}) \le \alpha\Delta(x^\rho_i | x^\rho_{1:i-1}) - \Delta(x^{k-m}_i | x^\rho_{1:i-1}) \label{eq:posub25}\\
		\le & \alpha \Delta(x^\rho_i | x^\rho_{1:i-1}) - \frac{k-m}{k} \Delta(x^\rho_i | x^\rho_{1:i-1}) \label{eq:posub26}\\
		=& (\alpha - 1 + m/k) \Delta(x^\rho_i | x^\rho_{1:i-1}), \label{eq:posub27}
	\end{align}
\end{subequations}
where \eqref{eq:posub22} is true by \eqref{eq:lemub_rho}, \eqref{eq:posub24} is true by submodularity of $f$. We conclude that when $m \le k$, $\beta = (\alpha -1 +m/k)$, implying that for this case
\begin{equation*}
	f(x^\pi) \le (2 \alpha + m/k) f(x^\rho).
\end{equation*}

Next suppose that $m \ge k$. Observe that $|\tilde{z}^\pi_i| \le k(n-1)$, since using the approximated augmented greedy policy, no more than $k(n-1)$ elements of $z^\pi_i$ can be chosen by other agents. This implies the following:
\begin{subequations}
	\begin{align}
		\Delta(& \tilde{z}_i^\pi | x^\rho_{1:i}) \le (1/\min(k/|\tilde{z}_i^\pi|, 1)) \cdot \max_{z \in (\tilde{z}^\pi_i)^k}\Delta(z | x^\rho_{1:i}) \\
		\le & \max(|\tilde{z}_i^\pi|/k, 1) \cdot \max_{z \in (\tilde{z}^\pi_i)^k}\Delta(z | x^\rho_{1:i}) \\
		\le & \max(\min(k(n-1), m)/k, 1) \cdot \max_{z \in (\tilde{z}^\pi_i)^k}\Delta(z | x^\rho_{1:i})  \\
		=& \min(n-1, m/k) \cdot \max_{z \in (\tilde{z}^\pi_i)^k}\Delta(z | x^\rho_{1:i})  \\
		\le & \alpha \cdot \min(n-1, m/k) \Delta(x^\rho_i | x^\rho_{1:i-1})
	\end{align}
\end{subequations}
We conclude that when $m \ge k$, $\beta = \alpha \cdot \min(n-1, m/k)$, implying that for this case:
\begin{equation*}
	f(x^\pi) \le (1 + \alpha (1 + \min(n-1, m/k))) f(x^\rho).
\end{equation*}

\subsection{Proof for Lemma \ref{lem:lb}} \label{app:lemlb}

We begin with
\begin{subequations}
	\begin{align}
		f(& x^\rho) \le f(x^\pi_{1:n-1}) + \Delta(x^\rho | x^\pi_{1:n-1}) \label{eq:lb1} \\
		= & f(x^\pi_{1:n-1}) + \Delta(x^\rho_n | x^\rho_{1:n-1}, x^\pi_{1:n-1}) \nonumber \\
		& + \sum_{i=1}^{n-1} \Delta(x^\rho_i | x^\rho_{1:i-1}, x^\pi_{1:n-1}) \label{eq:lb2} \\
		\le & f(x^\pi_{1:n-1}) + \Delta(x^\rho_n | x^\pi_{1:n-1}) + \sum_{i=1}^{n-1} \Delta(x^\rho_i | x^\pi_{1:i}) \label{eq:lb3} \\
		\le & f(x^\pi_{1:n-1}) + \alpha_1\Delta(x^\pi_n | x^\pi_{1:n-1}) + \sum_{i=1}^{n-1} \Delta(x^\rho_i | x^\pi_{1:i}) \label{eq:lb8}
	\end{align}
\end{subequations}
where \eqref{eq:lb1} and \eqref{eq:lb3} follow from submodularity of $f$, \eqref{eq:lb2} follows from the definition of $\Delta(\cdot)$, and \eqref{eq:lb8} follows from \eqref{eq:alpha1}. Focusing on the sum in \eqref{eq:lb8}, for any $0 \le \varepsilon_1, \dots \varepsilon_{n-1} \le 1$ (and defining ${\varepsilon_0 = 0}$), we see that
\begin{subequations}
	\begin{align}
		\sum_{i=1}^{n-1} & \Delta(x^\rho_i | x^\pi_{1:i}) = \sum_{i=1}^{n-1} (1 - \varepsilon_i) \Delta(x^\rho_i | x^\pi_{1:i})   + \sum_{i=1}^{n-1} \varepsilon_i \Delta(x^\rho_i | x^\pi_{1:i}) \label{eq:lb4} \\
		\le & \sum_{i=1}^{n-1} (1 - \varepsilon_i) \Delta(x^\rho_i | x^\pi_{1:i-1}) + \sum_{i=1}^{n-1} \alpha_2 \varepsilon_i \Delta^k(z^\pi_i | x^\pi_{1:i}) \label{eq:lb5} \\
		\le & \sum_{i=1}^{n-1} \alpha_1 (1 - \varepsilon_i) \Delta(x^\pi_i | x^\pi_{1:i-1}) + \sum_{i=1}^{n-1} \alpha_1\alpha_2 \varepsilon_i \Delta(x^\pi_{i+1} | x^\pi_{1:i}) \label{eq:lb6} \\
		= & \alpha_1\alpha_2\varepsilon_{n-1}\Delta(x^\pi_n | x^\pi_{1:n-1}) \nonumber \\
		& + \sum_{i=1}^{n-1}\alpha_1 (1 - \varepsilon_i + \alpha_2 \varepsilon_{i-1}) \Delta(x^\pi_i | x^\pi_{1:i-1}) \label{eq:lb7},
	\end{align}
\end{subequations}
where \eqref{eq:lb5} is true by submodularity of $f$ (1st term) and \eqref{eq:alpha2} (2nd term), \eqref{eq:lb6} is true by \eqref{eq:alpha1}, and \eqref{eq:lb7} is just a rearrangement of the terms. Applying this to \eqref{eq:lb8} yields
\begin{subequations}
	\begin{align}
		f(& x^\rho) \le f(x^\pi_{1:n-1}) + (\alpha_1 + \alpha_1\alpha_2\varepsilon_{n-1}) \Delta(x^\pi_n | x^\pi_{1:n-1}) \nonumber \\ 
		& + \sum_{i = 1}^{n-1} (\alpha_1 - \alpha_1\varepsilon_i + \alpha_1\alpha_2\varepsilon_{i-1}) \Delta(x^\pi_i | x^\pi_{1:i-1}) \\
		=& (\alpha_1 + \alpha_1\alpha_2\varepsilon_{n-1}) \Delta(x^\pi_n | x^\pi_{1:n-1}) \nonumber \\
		& + \sum_{i = 1}^{n-1} (1 + \alpha_1 - \alpha_1\varepsilon_i + \alpha_1\alpha_2\varepsilon_{i-1}) \Delta(x^\pi_i | x^\pi_{1:i-1}). \label{eq:lb9}
	\end{align}
\end{subequations}
Suppose that for a particular choice of $\varepsilon_i$, we let
\begin{equation}
	\varepsilon_i = \frac{ (1/\alpha_1)\sum_{j=0}^{i-1}\alpha_2^j}{\sum_{j=0}^{n-1}\alpha_2^j }.
\end{equation}
Since $\alpha_1, \alpha_2 \geq 1$, this satisfies the requirement that ${0 \leq \varepsilon_i \leq 1}$ for $i \in \{1, \dots, n-1\}$. Then
\begin{align}
	-\alpha_1 \varepsilon_i + \alpha_1\alpha_2 \varepsilon_{i-1} = & - \frac{ \sum_{j=0}^{i-1}\alpha_2^j}{\sum_{j=0}^{n-1}\alpha_2^j} + \frac{\sum_{j=1}^{i-1}\alpha_2^j}{\sum_{j=0}^{n-1}\alpha_2^j } \\
	=& -\frac{1}{\sum_{j=0}^{n-1}\alpha_2^j } \label{eq:lbfirst}
\end{align}
Likewise 
\begin{align}
	\alpha_1\alpha_2 \varepsilon_{n-1} = \frac{ \alpha_2\sum_{j=0}^{n-2}\alpha_2^j}{\sum_{j=0}^{n-1}\alpha_2^j }
	=&\frac{ \sum_{j=1}^{n-1}\alpha_2^j}{\sum_{j=0}^{n-1}\alpha_2^j } \\
	=& 1 - \frac{1}{\sum_{j=0}^{n-1}\alpha_2^j } \label{eq:lblast}
\end{align}
Applying \eqref{eq:lbfirst} and \eqref{eq:lblast} to \eqref{eq:lb9} yields
\begin{align}
	f(& x^\rho) \le \left(\alpha_1 + 1 -\frac{1}{\sum_{j=0}^{n-1}\alpha_2^j } \right) \Delta(x^\pi_n | x^\pi_{1:n-1}) \nonumber \\
	& + \sum_{i = 1}^{n-1} \left(1 + \alpha_1 -\frac{1}{\sum_{j=0}^{n-1}\alpha_2^j } \right) \Delta(x^\pi_i | x^\pi_{1:i-1}) \\
	=& \left(1 + \alpha_1 -\frac{1}{\sum_{j=0}^{n-1}\alpha_2^j } \right) f(x^\pi)
\end{align}

\begin{IEEEbiography}[{\includegraphics[width=1in,height=1.25in,clip,keepaspectratio]{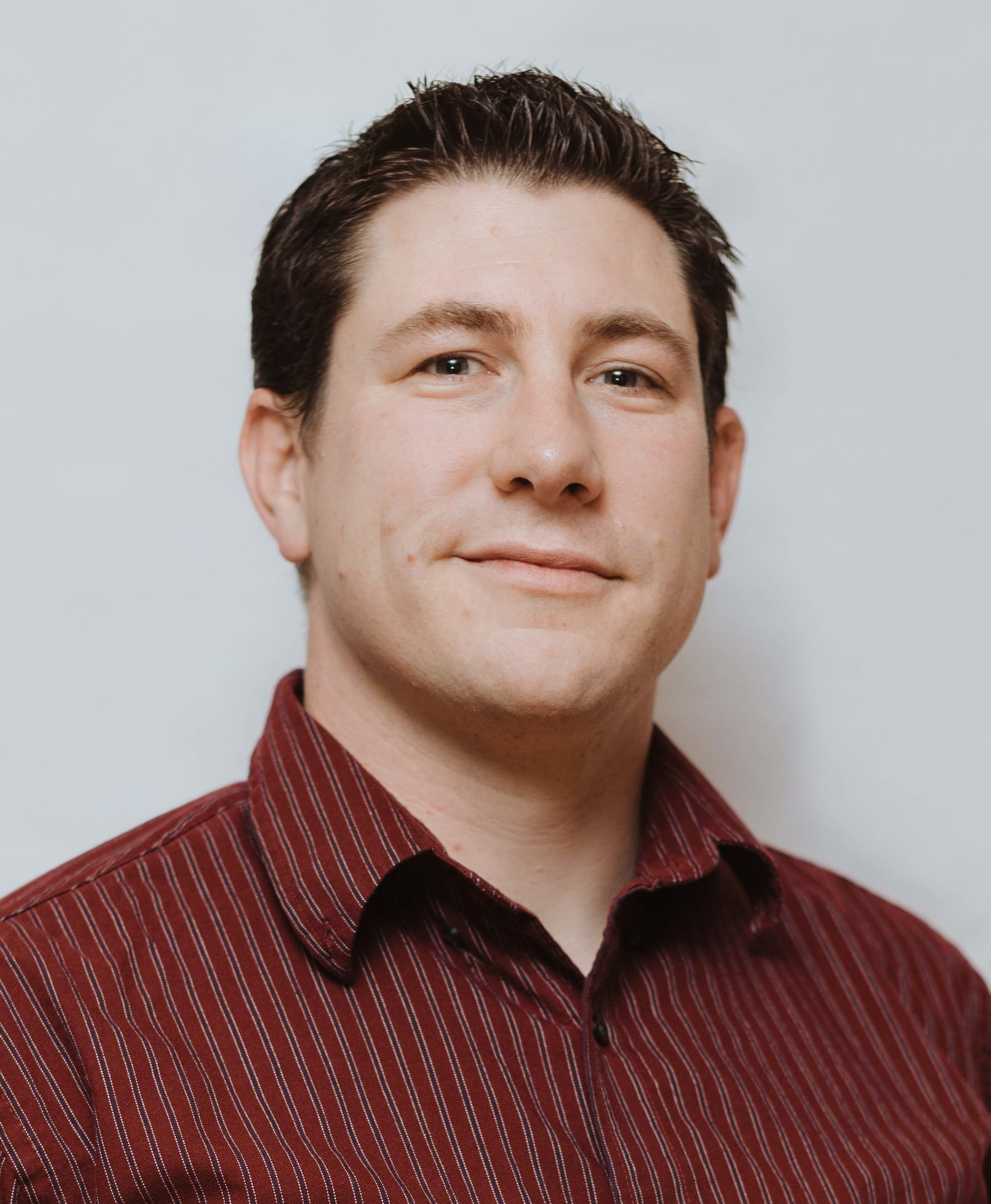}}]
	{David Grimsman} is an Assistant Professor in the Computer Science Department at Brigham Young University. He completed BS in Electrical and Computer Engineering at Brigham Young University in 2006 as a Heritage Scholar, and with a focus on signals and systems. After working for BrainStorm, Inc. for several years as a trainer and IT manager, he returned to Brigham Young University and earned an MS in Computer Science in 2016. He then received his PhD in Electrical and Computer Engineering from UC Santa Barbara in 2021. His research interests include mulit-agent systems, game theory, distributed optimization, network science, linear systems theory, and security of cyberphysical systems.
\end{IEEEbiography}
\begin{IEEEbiography}[{\includegraphics[width=1in,height=1.25in,clip,keepaspectratio]{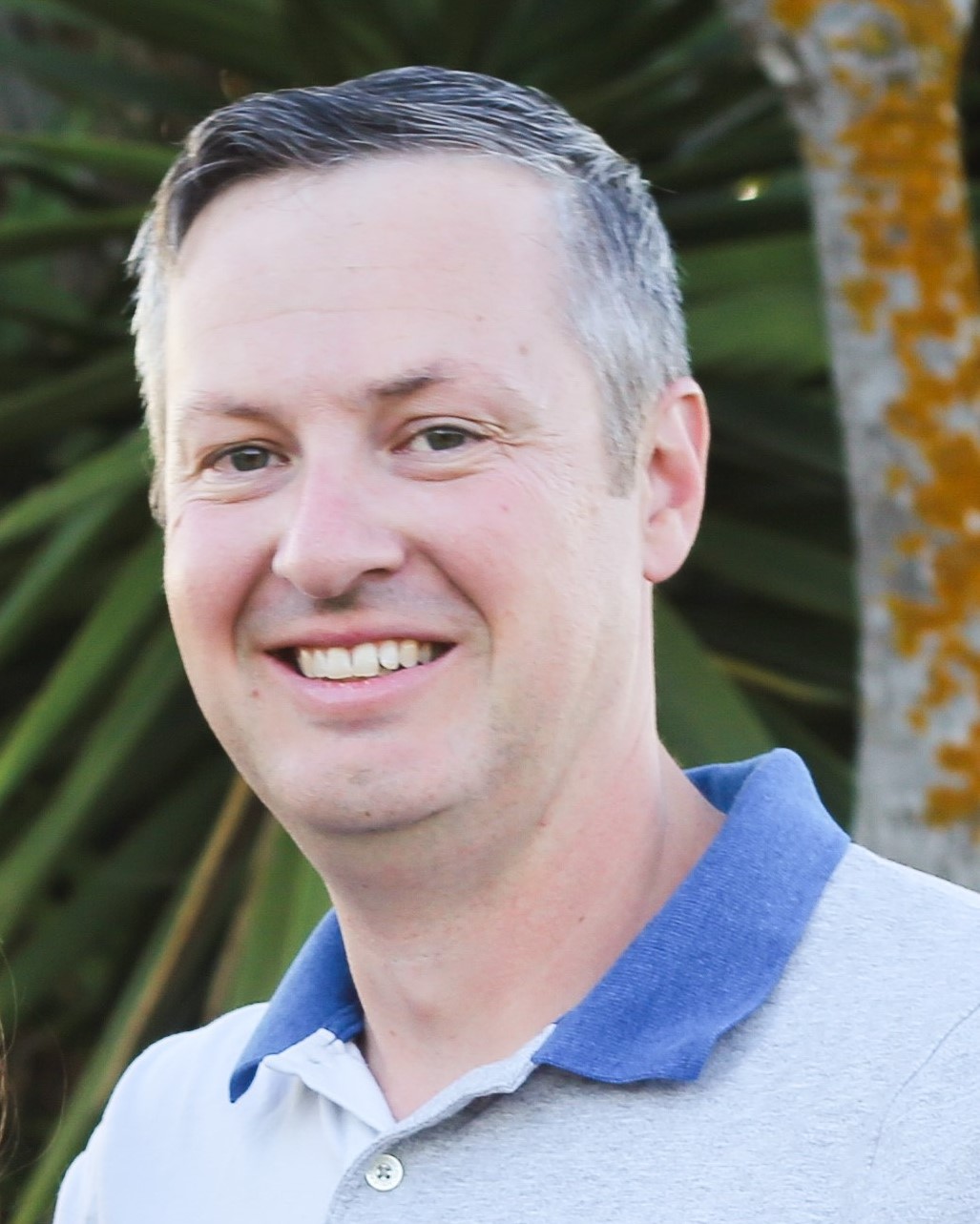}}]
{Matthew R. Kirchner} received his B.S. in Mechanical Engineering from Washington State University in 2007 and his M.S. in Electrical Engineering from the University of Colorado at Boulder in 2013. In 2007 he joined the Naval Air Warfare Center Weapons Division in the Navigation and Weapons Concepts Develop Branch and in 2012 transferred into the Image and Signal Processing Branch in the Research and Intelligence Department, Code D5J1000. He is currently a Ph.D. student in the Electrical and Computer Engineering Department at the University of California, Santa Barbara. His research interests include level set methods for optimal control, differential games, and reachability; multi-vehicle robotics; nonparametric signal and image processing; and navigation and flight control. He was the recipient of a Naval Air
Warfare Center Weapons Division Graduate Academic Fellowship from 2010 to 2012 and in 2011 was
named a Paul Harris Fellow by Rotary International. Matthew is a student member of the IEEE.
\end{IEEEbiography}
\begin{IEEEbiography}[{\includegraphics[width=1in,height=1.25in,clip,keepaspectratio]{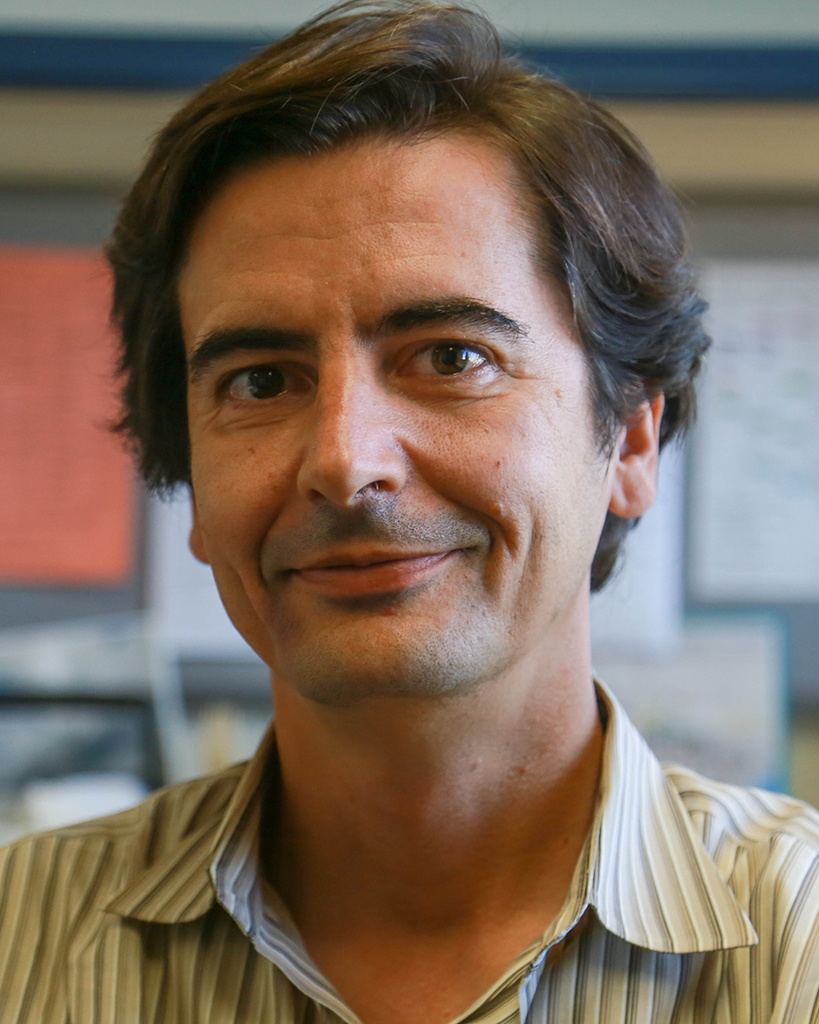}}]%
{Jo\~{a}o P. Hespanha} was born in Coimbra, Portugal, in 1968. He received the Licenciatura in electrical and computer engineering from the Instituto Superior Técnico, Lisbon, Portugal in 1991 and the Ph.D. degree in electrical engineering and applied science from Yale University, New Haven, Connecticut in 1998. From 1999 to 2001, he was Assistant Professor at the University of Southern California, Los Angeles. He moved to the University of California, Santa Barbara in 2002, where he currently holds a Professor position with the Department of Electrical and Computer Engineering.

Dr. Hespanha is the recipient of the Yale University’s Henry Prentiss Becton Graduate Prize for exceptional achievement in research in Engineering and Applied Science, a National Science Foundation CAREER Award, the 2005 best paper award at the 2nd Int. Conf. on Intelligent Sensing and Information Processing, the 2005 Automatica Theory/Methodology best paper prize, the 2006 George S. Axelby Outstanding Paper Award, and the 2009 Ruberti Young Researcher Prize. Dr. Hespanha is a Fellow of the International Federation of Automatic Control (IFAC) and of the IEEE. He was an IEEE distinguished lecturer from 2007 to 2013.

His current research interests include hybrid and switched systems; multi-agent control systems; game theory; optimization; distributed control over communication networks (also known as networked control systems); the use of vision in feedback control; stochastic modeling in biology; and network security.
\end{IEEEbiography}
\begin{IEEEbiography}[{\includegraphics[width=1in,height=1.25in,clip,keepaspectratio]{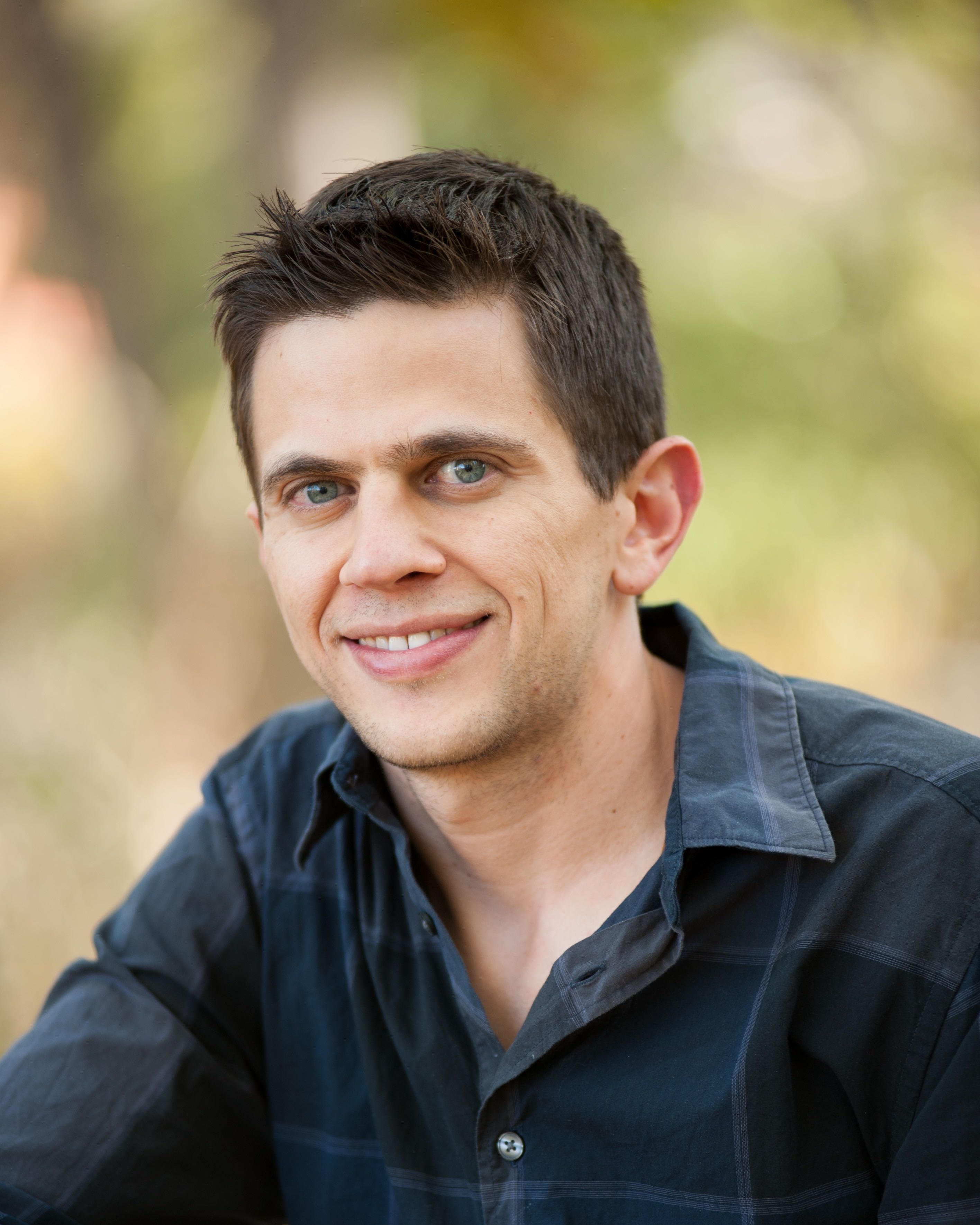}}]
{Jason R. Marden} is an associate professor in the Department of Electrical and Computer Engineering at the University of California, Santa Barbara. He received the B.S. degree in 2001 and the Ph.D. degree in 2007 (under the supervision of Jeff S. Shamma), both in mechanical engineering from the University of California, Los Angeles, where he was awarded the Outstanding Graduating Ph.D. Student in Mechanical Engineering. After graduating, he was a junior fellow in the Social and Information Sciences Laboratory at the California Institute of  Technology until 2010 and then an assistant professor at the University of Colorado until 2015. He is a recipient of an ONR Young Investigator Award (2015), an NSF Career Award (2014), the AFOSR Young Investigator Award (2012), the SIAM CST Best Sicon Paper Award (2015), and the American Automatic Control Council Donald P. Eckman Award (2012). His research interests focus on game-theoretic methods for the control of distributed multiagent systems.
\end{IEEEbiography}

\end{document}